\documentclass[aps,twocolumn,superscriptaddress,showpacs,amssymb,prl,floatfix,longbibliography]{revtex4-1}

\usepackage{graphicx,color}
\usepackage{dcolumn}
\usepackage{amsmath}
\usepackage{amssymb}
\usepackage{epstopdf}

\usepackage[]{hyperref}
\hypersetup{colorlinks=true,linkcolor=blue,citecolor=blue,urlcolor=blue,pdfpagemode=UseNone}

\begin{document}

\thispagestyle{myheadings}

\title{Uniaxial strain control of spin-polarization in multicomponent nematic
order of BaFe$_{2}$As$_{2}$}

\author{T. Kissikov}

\affiliation{Department of Physics, University of California, Davis, California
95616, USA}

\author{R. Sarkar}

\affiliation{Institute for Solid State Physics, TU Dresden, D-01069 Dresden, Germany}

\author{M. Lawson}

\author{B. T. Bush}

\affiliation{Department of Physics, University of California, Davis, California
95616, USA}

\author{E. I. Timmons}

\author{M. A. Tanatar}

\author{R. Prozorov}

\author{S. L. Bud'ko}

\author{P. C. Canfield}

\affiliation{Ames Laboratory U.S. DOE and Department of Physics and Astronomy,Iowa
State University, Ames, Iowa 50011, USA}

\author{R. M. Fernandes}

\affiliation{School of Physics and Astronomy, University of Minnesota, Minneapolis,
Minnesota 55455, USA}

\author{N. J. Curro}

\affiliation{Department of Physics, University of California, Davis, California
95616, USA}

\date{\today}
\begin{abstract}
\textbf{The iron-based high temperature superconductors exhibit a
rich phase diagram reflecting a complex interplay between spin, lattice,
and orbital degrees of freedom \cite{doping122review,FernandesNematicPnictides,Fernandes2012,FernandesPRLnematicT1}.
The nematic state observed in many of these compounds epitomizes this
complexity, by entangling a real-space anisotropy in the spin fluctuation
spectrum with ferro-orbital order and an orthorhombic lattice distortion
\cite{TanatarTensileStressPRB,StrainedPnictidesNS2014science,StrainBa122neutronsPRB2015}.
A more subtle and much less explored facet of the interplay between
these degrees of freedom arises from the sizable spin-orbit coupling
present in these systems, which translates anisotropies in real space
into anisotropies in spin space. Here, we present a new technique
enabling nuclear magnetic resonance under precise tunable strain control,
which reveals that upon application of a tetragonal symmetry-breaking
strain field, the magnetic fluctuation spectrum in the paramagnetic phase
of BaFe$_{2}$As$_{2}$ also acquires an anisotropic response in spin-space.
Our results unveil a hitherto uncharted internal spin structure of the nematic order parameter, indicating that similar to liquid crystals, electronic nematic materials may offer a novel route to
magneto-mechanical control. }
\end{abstract}
\maketitle
In the absence of external strain, BaFe$_{2}$As$_{2}$ undergoes
a weakly first-order antiferromagnetic phase transition at $T_{N}=135$K,
accompanied by an orthorhombic structural distortion that breaks the
tetragonal symmetry of the unit cell in the paramagnetic phase. The
relatively small orthorhombic lattice distortion ($\sim0.3\%$) \cite{TanatarTensileStressPRB}
is driven by a nematic instability \cite{FradkinNematicReview}, whose
electronic origin is manifested by the large in-plane resistivity
anisotropy ($\sim100\%$) \cite{TanatarDetwin,IronArsenideDetwinnedFisherScience2010}.
Despite being simultaneous in BaFe$_{2}$As$_{2}$, the nematic and
antiferromagnetic transition temperatures, $T_{s}$ and $T_{N}$,
split upon doping, giving rise to a regime with long-range nematic
order but no antiferromagnetic order, since $T_{N}<T_{s}$ \cite{doping122review,FernandesSchmalianNatPhys2014}.

\begin{figure}[!tb]
\centering \includegraphics[width=1\linewidth]{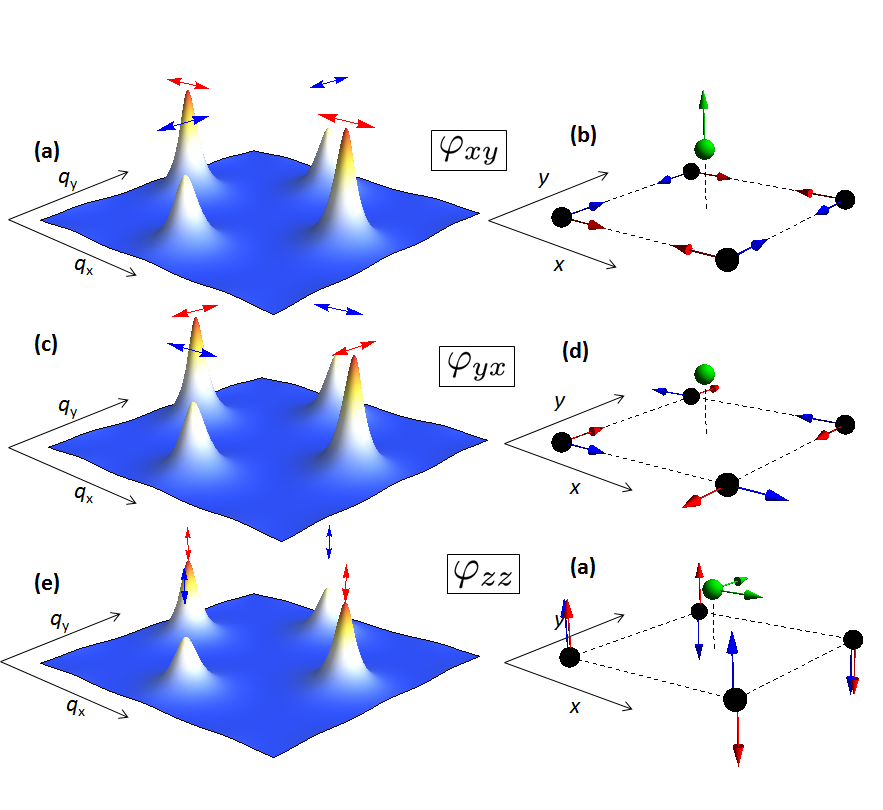}
\caption{\label{fig:multinematic} Spin fluctuations in momentum space (left)
and in real space (right) and polarization directions of the Fe spins
for the three nematic components, $\varphi_{xy}$ (a,b), $\varphi_{yx}$
(c,d), and $\varphi_{zz}$ (e,f). The red arrows correspond to the
magnetic ordering vector $\mathbf{Q}_{1}=(\pi,0)$ and the blue arrows
correspond to $\mathbf{Q}_{2}=(0,\pi)$. The black spheres are the
Fe sites, the green sphere is the As site, and the green arrows indicate
the direction of the hyperfine field.}
\end{figure}

The close relationship between nematicity and the magnetic degrees
of freedom can be seen directly from the stripe-like nature of the
antiferromagnetic state, which orders with one of two possible wave-vectors
related by a $90^{\circ}$ rotation: $\mathbf{Q}_{1}=(\pi,0)$ (corresponding
to spins parallel along the $y$ axis and anti-parallel along $x$)
and $\mathbf{Q}_{2}=(0,\pi)$ (corresponding to spins parallel along
$x$ and anti-parallel along $y$). Below $T_{N}$ nearest neighbor
spins are parallel or antiparallel depending on whether they are connected
by a short or long bond, however above $T_{N}$ but below $T_{s}$ the magnetic
fluctuations centered around $\mathbf{Q}_{1}$ become weaker or stronger
than those centered around $\mathbf{Q}_{2}$,
depending on whether the $b$ axis is parallel or perpendicular to
$\mathbf{Q}_{1}$, respectively. Mathematically, this allows one to
define the nematic order parameter $\bar{\varphi}$ in terms of the
(spin unpolarized) magnetic susceptibility $\chi\left(\mathbf{q}\right)$
according to $\bar{\varphi}\equiv\chi^{-1}(\mathbf{Q}_{2})-\chi^{-1}(\mathbf{Q}_{1})$
\cite{FernandesNematicPnictides}. Such an interplay between nematic
and spin degrees of freedom has been indeed observed by neutron scattering
\cite{DaiRMP2015,StrainBa122neutronsPRB2015,StrainedPnictidesNS2014science,PengchangPRB2015}
and nuclear magnetic resonance (NMR) \cite{DioguardiNematicGlass2015,NMRnematicStrainBa122PRB2016,PhysRevB.89.214511}
experiments in detwinned BaFe$_{2}$As$_{2}$ crystals.

However, orbital degrees of freedom also participate actively in the
nematic phase. This leads to the well known effect that tetragonal
symmetry-breaking is also manifested by a ferro-orbital polarization
that makes the occupation of the Fe $d_{xz}$ orbitals different than
the occupation of the Fe $d_{yz}$ orbitals. A less explored effect
emerges from the relatively sizable spin-orbit coupling (SOC), which
converts anisotropies in real space into anisotropies in spin space.
On one hand,  SOC enforces the spins to point along
the ordering vector direction below $T_{N}$. On the other hand,  SOC leads to different magnitudes of
the diagonal spin susceptibility components, $\chi_{\alpha\alpha}\left(\mathbf{q}\right)$
with $\alpha=(x,y,z)$, in the nematic temperature
regime, $T_{N}<T<T_{s}$. As a result, the nematic order parameter naturally
acquires an internal spin structure, since generically one must define
$\varphi_{\alpha\beta}=\chi_{\alpha\alpha}^{-1}(\mathbf{Q}_{2})-\chi_{\beta\beta}^{-1}(\mathbf{Q}_{1})$.
Clearly, the nematic order parameter $\bar{\varphi}$ defined above
can be understood as an average over all possible polarizations, $\bar{\varphi}=\frac{1}{9}\sum\limits _{\alpha\beta}\varphi_{\alpha\beta}$.
The space-group symmetry of the iron pnictides enforces many of these
combinations to vanish, yielding only three non-zero independent components:
$\varphi_{xy}$, $\varphi_{yx}$, and $\varphi_{zz}$. The physical
meaning of each component is depicted in Fig. \ref{fig:multinematic};
for instance, $\varphi_{xy}$ is a measure of the asymmetry between
spin fluctuations peaked at $\mathbf{Q}_{1}$ and polarized along
the $x$ axis, and spin fluctuations peaked at $\mathbf{Q}_{2}$ and
polarized along the $y$ axis.
\begin{figure}[!tb]
\centering \includegraphics[width=\linewidth]{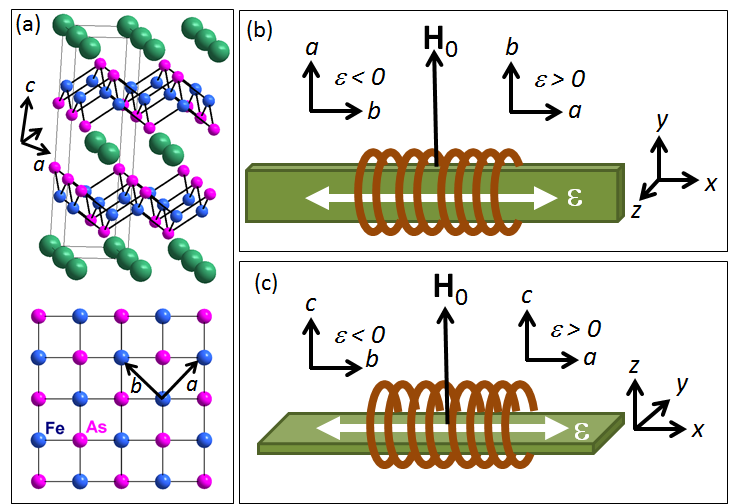} 
\caption{\label{fig:cartoon} (a) Crystal structure of BaFe$_{2}$As$_{2}$,
with Ba (green), Fe (blue) and As (magenta) sites shown. Lower panel
shows the Fe-As plane in the tetragonal phase, with arrows indicating
the unit cell axes of the orthorhombic phase ($\mathbf{a}~||~(110)_{tet}$,
$\mathbf{b}~||~(1\overline{1}0)_{tet}$). (b,c) Orientation of the
magnetic field with respect to the coil ($H_{1}$) and strain axis
for $\mathbf{H}_{0}\perp\mathbf{c}$ (b) and $\mathbf{H}_{0}~||~\mathbf{c}$
(c). For positive (tensile) strain $\mathbf{H}_{0}$ is parallel to
$\mathbf{b}$, whereas for negative (compressive) strain $\mathbf{H}_{0}$
is along $\mathbf{a}$.}
\end{figure}

Elucidating the hitherto unkown spin structure of the nematic order
parameter is fundamental to shed light on the intricate interplay
between orbital, spin, and lattice degrees of freedom, which are ultimately
responsible for the superconducting instability of the system. In
this paper, we perform NMR spin-lattice relaxation measurements to
probe the anisotropy of the spin fluctuations under fixed strain in
the paramagnetic phase of BaFe$_{2}$As$_{2}$. The role of the applied
uniaxial strain is to provide a small tetragonal symmetry-breaking
field, akin to externally applied magnetic fields in ferromagnets.
In contrast to previous works, here we probe the magnetic fluctuations
anisotropy both in real space and in spin space \textendash{} more
specifically, we determine each of the nematic susceptibilities associated
with the three nematic components $\varphi_{xy}$, $\varphi_{yx}$,
and $\varphi_{zz}$. This is possible because the magnetic fluctuations
associated with each spin polarization pattern generate very different
types of fluctuating local fields experienced by the $^{75}$As nuclear
spin ($I=3/2$), which couples to the four nearest neighbor Fe spins
via a transferred hyperfine interaction (see Fig. \ref{fig:multinematic})
\cite{T1formfactorsArsenides}. Our main result is that the three
nematic components respond differently to external strain, i.e. nematic
order induces not only real-space anisotropy, but also affects the
spin-space anisotropy. In particular, we find that the out-of-plane
spin fluctuations centered at $\mathbf{Q}\parallel\hat{a}$ are more
strongly enhanced by the strain, as compared to the spin fluctuations
polarized along the longer in-plane axis. This raises the interesting
possibility of reversing the spin polarization of the system from
in-plane to out-of-plane by applying a sufficiently strong in-plane
strain. More broadly, our results thus opens a new avenue toward magneto-mechanical
manipulation of strongly correlated systems that display nematic order.

Key to this study is our ability to control precisely the uniaxial
strain applied in the sample, which is achieved by integrating a novel
piezoelectric strain cell with an NMR probe. This new device is based
upon a design used previously to investigate the superconducting transition
temperature of Sr$_{2}$RuO$_{4}$ \cite{Hicks2014,Sr2RuO4strainScience2014},
and can achieve both positive and negative strains with large
strain homogeneity. This device differs from the horseshoe-clamp \cite{TanatarDetwin}
used previously for NMR \cite{NMRnematicStrainBa122PRB2016}, and
offers superior control over the sample alignment and the level of
strain applied.

\begin{figure*}[!tb]
\centering
\includegraphics[width=0.6\linewidth]{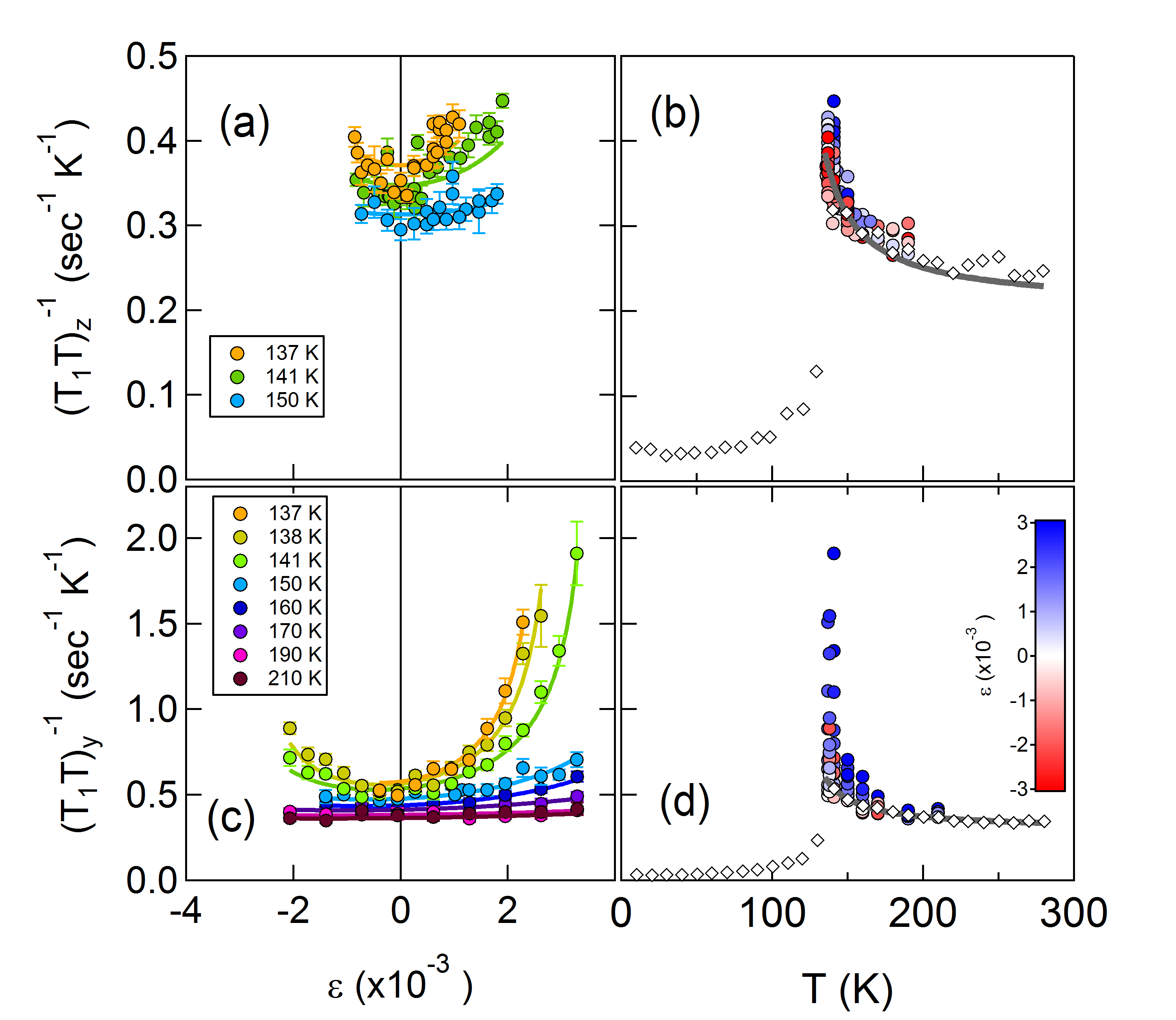}
\includegraphics[width=0.39\linewidth]{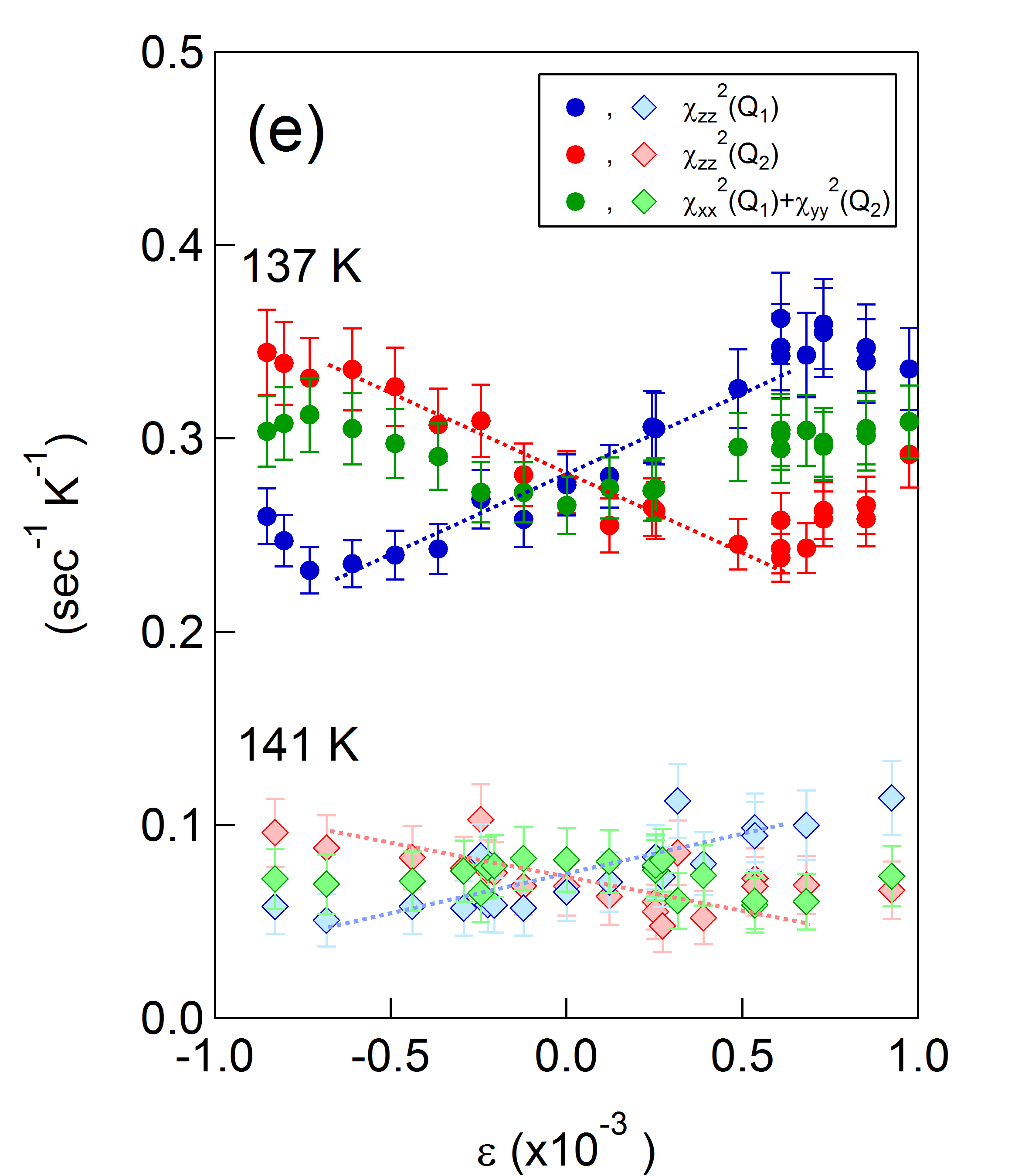}
\caption{\label{fig:T1Tinv} $(T_{1}T)_{y,z}^{-1}$ versus strain (a,c) and
versus temperature (b,d) The solid lines are fits as described in
the text. The open diamonds in (b,d) are reproduced from Ref. \protect\onlinecite{takigawa2008}.
(e) $\chi_{zz}\left(\mathbf{Q}_{1}\right)$, $\chi_{zz}\left(\mathbf{Q}_{2}\right)$, and $\chi_{xx}\left(\mathbf{Q}_{1}\right) + \chi_{yy}\left(\mathbf{Q}_{2}\right)$ as a function
of strain at 137K and 141K. The data have been displaced vertically
for clarity. The dashed lines are guides to the eye.}
\end{figure*}

Single crystals of BaFe$_{2}$As$_{2}$ were cut along the tetragonal
(110) direction and mounted in the cryogenic strain cell with field
oriented both parallel and perpendicular to the crystallographic $c$-axis,
as shown in Fig. \ref{fig:cartoon}. The strain cell contains two
sets of piezoelectric stacks, one inner and two outer. Because the
sample is freely suspended between the piezoelectric stacks rather
than glued down over a portion of the stack, the full displacement
of each stack is transferred to the sample. As a result the device
is able to achieve displacements of $\pm6\mu$m at room temperature
and $\pm3\mu$m at 4K, corresponding to strains of the order of $10^{-3}$ in this material.
A free-standing NMR coil was placed around the sample prior to securing
the ends of the crystal in the strain device with epoxy. The radiofrequency
field $\mathbf{H}_{1}$ is oriented parallel to the strain axis, which
is always perpendicular to the external field, $\mathbf{H}_{0}$.
In our device, strain is always applied along the $x$ axis defined
in Fig. \ref{fig:cartoon}; since the $b$ axis is defined as the
shorter axis, positive (i.e. tensile) strain corresponds to $x\parallel a$
and $y\parallel b$, whereas negative (i.e. compressive) strain gives
$y\parallel a$ and $x\parallel b$. When the crystal is strained
by applying voltage to the piezoelectric stacks, the displacement,
$x$, is measured by a capacitive position sensor, and strain is calculated
as $\epsilon=(x-x_{0})/L_{0}$, where $L_{0}$ is the unstrained length
of the crystal. To account for differential thermal contraction, the zero-strain displacement, $x_{0}$, was determined
by the condition that the quadrupolar splitting $\nu_{\alpha\alpha}$
satisfies the tetragonal-symmetry relationship $|\nu_{xx}|=|\nu_{yy}|=|\nu_{zz}|/2$,
as described in the supplemental material. The linear relationship between  $\nu_{\alpha\alpha}$ and strain (Fig. S1) indicates
that both positive and negative strains are achieved, without bowing of the crystal. The field $\mathbf{H}_{0}$
was oriented either along the $z$ direction parallel to the crystalline
$c$ axis, or in the plane of the crystal along the $y$-direction,
as shown in Fig. \ref{fig:cartoon}.

The spin lattice relaxation rate $(T_{1}T)_{\mu}^{-1}$ for different
field orientations $\mu=z,\,y$ is shown in Fig. \ref{fig:T1Tinv}
both as a function of strain $\varepsilon$ and temperature $T$.
It is striking that while $(T_{1}T)_{z}^{-1}$ increases by approximately
30\% at 137K for the largest applied strain (approximatelly $0.3\%$),
$(T_{1}T)_{y}^{-1}$ increases by 500\%. In both cases, both positive
and negative strain increase $(T_{1}T)^{-1}$ in a nonlinear fashion.
This behavior is a manifestation of the spin anisotropy induced by
nematic order. More precisely, the spin lattice relaxation rate is
primarily dominated by the fluctuations of the local hyperfine field
at the As site, which in turn is determined by the neighboring iron
spins according to:
\begin{equation}
\left(\frac{1}{T_{1}T}\right)_{\mu}=\frac{\gamma^{2}}{2}\lim_{\omega\rightarrow0}\sum\limits _{\mathbf{q},\alpha,\beta}\mathcal{F}_{\alpha\beta}^{(\mu)}(\mathbf{q})\frac{\textrm{Im}\chi_{\alpha\beta}(\mathbf{q},\omega)}{\hslash\omega},\label{eqn:dynamical_susceptibility}
\end{equation}
where $\gamma$ is the nuclear gyromagnetic factor, $\mathcal{F}_{\alpha\beta}^{(\mu)}$
are the hyperfine form factors, which depend on the field direction
$\mu$ (see Supplemental Material), $\chi_{\alpha\beta}(\mathbf{q},\omega)$
is the dynamical magnetic susceptibility, and $\alpha,\beta=\left\{ x,y,z\right\} $
\cite{T1formfactorsArsenides}. Because the system is metallic, spin
fluctuations experience Landau damping, resulting in the low-energy
dynamics $\chi_{\alpha\beta}^{-1}(\mathbf{q},\omega)=\chi_{\alpha\beta}^{-1}(\mathbf{q})-i\hbar\omega/\Gamma$,
where $\Gamma$ is the Landau damping, as seen by neutron scattering
experiments. Consequently, $\lim\limits _{\omega\rightarrow0}\frac{\textrm{Im}\chi_{\alpha\beta}(\mathbf{q},\omega)}{\hslash\omega}=\frac{1}{\Gamma}\chi_{\alpha\beta}^{2}(\mathbf{q})$,
i.e. the spin-lattice relaxation rate is proportional to the squared
susceptibility integrated over the entire Brillouin zone. Since the
magnetically ordered state has wave-vectors $\mathbf{Q}_{1}=\left(\pi,0\right)$
and $\mathbf{Q}_{2}=\left(0,\pi\right)$, one expects that the susceptibility
is peaked at these two momenta. Indeed, neutron scattering experiments
confirm that the magnetic spectral weight is strongly peaked at $\mathbf{Q}_{1}$
and $\mathbf{Q}_{2}$.

Therefore, as an initial step to elucidate the effect of strain on
the spin fluctuations anisotropy, we consider that the susceptibility
is sharply peaked at these two magnetic ordering vectors. Evaluation
of the hyperfine form factors yields:

\begin{align}
\left(T_{1}T\right)_{x}^{-1} & \propto\chi_{xx}^{2}\left(\mathbf{Q}_{1}\right)+\chi_{yy}^{2}\left(\mathbf{Q}_{2}\right)+\chi_{zz}^{2}\left(\mathbf{Q}_{2}\right)\nonumber \\
\left(T_{1}T\right)_{y}^{-1} & \propto\chi_{xx}^{2}\left(\mathbf{Q}_{1}\right)+\chi_{yy}^{2}\left(\mathbf{Q}_{2}\right)+\chi_{zz}^{2}\left(\mathbf{Q}_{1}\right)\nonumber \\
\left(T_{1}T\right)_{z}^{-1} & \propto\chi_{zz}^{2}\left(\mathbf{Q}_{1}\right)+\chi_{zz}^{2}\left(\mathbf{Q}_{2}\right)\label{eq:T1T}
\end{align}
where the prefactors are approximately the same in all equations
(see SM), and proportional to the off-diagonal hyperfine matrix element
$\mathcal{F}_{xz}$ coupling in-plane Fe spin fluctuations to out-of-plane
As hyperfine fields (and vice-versa). The fact that $\chi_{zz}\left(\mathbf{Q}_{i}\right)$
contributes to $T_{1}$ for all directions of the applied magnetic
field is thus consistent with the hyperfine field analysis depicted
in Fig. \ref{fig:multinematic}, since out-of-plane spin fluctuations
on the Fe sites produce hyperfine fluctuating fields in the As sites
along both in-plane directions. Similarly, the fact that only $\chi_{xx}\left(\mathbf{Q}_{1}\right)$
and $\chi_{yy}\left(\mathbf{Q}_{2}\right)$ contribute to $T_{1}$
for external fields applied along the plane is a consequence of the
fact that these spin fluctuations generate hyperfine fields in the
As site oriented out of the plane.

Because by symmetry $\left(T_{1}T\right)_{x}^{-1}\left(\varepsilon\right)=\left(T_{1}T\right)_{y}^{-1}\left(-\varepsilon\right)$,
the NMR data can be used to extract the strain and temperature dependence
of the three polarized spin-susceptibility combinations $\chi_{zz}^{2}\left(\mathbf{Q}_{1}\right)$,
$\chi_{zz}^{2}\left(\mathbf{Q}_{2}\right)$, and $\chi_{xx}^{2}\left(\mathbf{Q}_{1}\right)+\chi_{yy}^{2}\left(\mathbf{Q}_{2}\right)$,
as shown in Fig. \ref{fig:T1Tinv}(e). This analysis provides several
interesting insights. First, focusing on the out-of-plane fluctuations,
in-plane strain enhances spin fluctuations around one of the two ordering
vectors ($\chi_{zz}\left(\mathbf{Q}_{1}\right)$ for $\varepsilon>0$
and $\chi_{zz}\left(\mathbf{Q}_{2}\right)$ for $\varepsilon<0$)
at the same time as it suppresses the fluctuations around the other
ordering vector. Therefore, in-plane strain transfers magnetic spectral
weight between the two dominant out-of-plane spin-fluctuation channels.
This is consistent with neutron scattering experiments in detwinned
pnictides \cite{StrainedPnictidesNS2014science}, which however only
probed the unpolarized susceptibility. More importantly, this behavior
is a direct manifestation of the response of the nematic order parameter
$\varphi_{zz}$ to strain, since $\varphi_{zz}=\chi_{zz}^{-1}(\mathbf{Q}_{2})-\chi_{zz}^{-1}(\mathbf{Q}_{1})$.

Turning now to the average in-plane fluctuations $\chi_{xx}^{2}\left(\mathbf{Q}_{1}\right)+\chi_{yy}^{2}\left(\mathbf{Q}_{2}\right)$,
we note that, in contrast to the quantity $\chi_{zz}\left(\mathbf{Q}_{1}\right)-\chi_{zz}\left(\mathbf{Q}_{2}\right)$,
it is an even function of the applied strain. This behavior can be
attributed to the response of the nematic order parameter $\varphi_{xy}=\chi_{xx}^{-1}(\mathbf{Q}_{2})-\chi_{yy}^{-1}(\mathbf{Q}_{1})$
to strain. Similarly to $\varphi_{zz}$, $\varphi_{xy}$ promotes
a transfer of magnetic spectral weight, but now between $x$-polarized
spin fluctuations around $\mathbf{Q}_{1}$ and $y$-polarized spin
fluctuations around $\mathbf{Q}_{2}$. Since only the combination
$\chi_{xx}^{2}\left(\mathbf{Q}_{1}\right)+\chi_{yy}^{2}\left(\mathbf{Q}_{2}\right)$
contributes to the spin-lattice relation rate, the total magnetic
spectral weight remains the same to linear order in $\varphi_{xy}$,
since what is suppressed in, say, $\chi_{yy}(\mathbf{Q}_{2})$ is
tranferred to $\chi_{xx}(\mathbf{Q}_{1})$. Of course, as strain is
enhanced, non-linear effects quadratic in $\varphi_{xy}^{2}$ take
place, in agreement with the behavior displayed by Fig. \ref{fig:T1Tinv}(e).
Note that the third nematic order parameter, $\varphi_{yx}=\chi_{yy}^{-1}(\mathbf{Q}_{2})-\chi_{xx}^{-1}(\mathbf{Q}_{1})$,
does not affect the in-plane fluctuations that contribute the most
to the spin-lattice relaxation rate. This is not unexpected, since
the spin fluctuations associated with $\chi_{yy}(\mathbf{Q}_{1})$
and $\chi_{xx}(\mathbf{Q}_{2})$ do not generate hyperfine fields
in the As sites, as shown in Fig. \ref{fig:multinematic}.

The most striking feature of Fig. \ref{fig:T1Tinv}(e) is that the
out-of-plane spin fluctuations seem to have a larger response to in-plane
strain than the in-plane spin fluctuations. This observation suggests
that the nematic susceptibility associated with $\varphi_{zz}$, $\chi_{\mathrm{nem}}^{(zz)}\equiv\partial\varphi_{zz}/\partial\varepsilon$,
is larger than the nematic susceptibility associated with $\varphi_{xy}$,
$\chi_{\mathrm{nem}}^{(xy)}\equiv\partial\varphi_{xy}/\partial\varepsilon$,
and is manifestation of the fact that nematic order induces not only
real-space anisotropy, but also spin-space anisotropy. To make this
analysis more quantitative, we fit the full temperature, strain, and
field orientation dependence of $T_{1}$ to a model that incorporates
the fact that the magnetic fluctuations are not infinitely peaked
at the ordering vectors $\mathbf{Q}_{1,2}$, since the magnetic correlation
length is finite above the magnetic transition. In the tetragonal
phase, there are three different magnetic correlation lengths, $\xi_{x}$,
$\xi_{y}$, and $\xi_{z}$, associated respectively with the pairs
of peaks $\left(\chi_{xx}\left(\mathbf{Q}_{1}\right),\chi_{yy}\left(\mathbf{Q}_{2}\right)\right)$;
$\left(\chi_{yy}\left(\mathbf{Q}_{1}\right),\chi_{xx}\left(\mathbf{Q}_{2}\right)\right)$,
and $\left(\chi_{zz}\left(\mathbf{Q}_{1}\right),\chi_{zz}\left(\mathbf{Q}_{2}\right)\right)$.
This spin anisotropy is intrinsic to the tetragonal crystalline symmetry
and is enforced by the spin-orbit coupling even in the absence of
nematic order. Nematic order induced by strain breaks the equivalence
between these pairs of peaks, splitting the correlation lengths into
$\tilde{\xi}_{x}^{-2}=\xi_{x}^{-2}\mp\varphi_{xy}$, $\tilde{\xi}_{y}^{-2}=\xi_{y}^{-2}\mp\varphi_{yx}$,
and $\tilde{\xi}_{z}^{-2}=\xi_{z}^{-2}\mp\varphi_{zz}$. This model
is similar to the one used previously in \cite{NMRnematicStrainBa122PRB2016}
and is described in the supplemental material.

\begin{figure}[!tb]
\centering \includegraphics[width=1\linewidth]{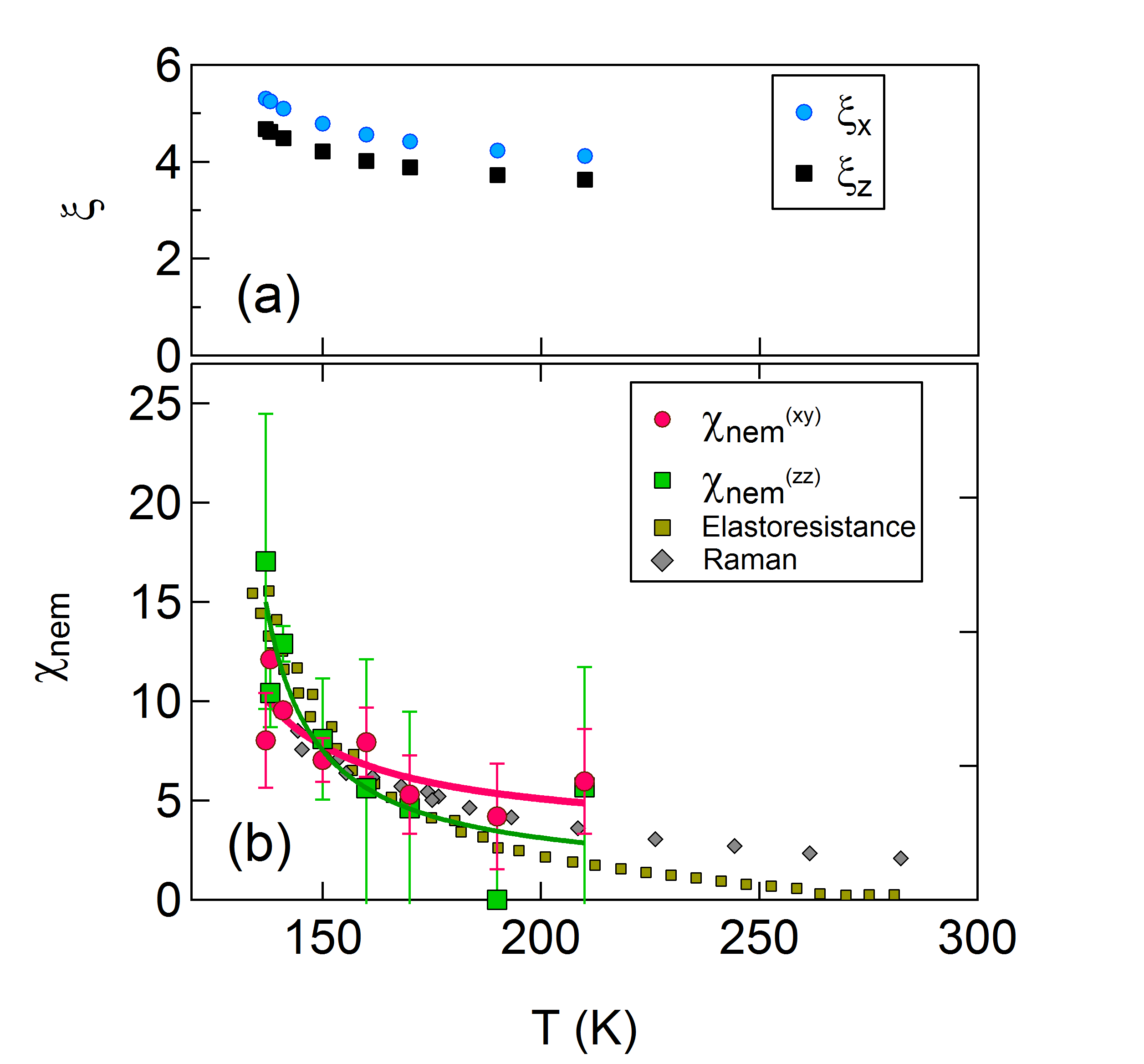}
\caption{\label{fig:fitpars} Fit parameters (a) $\kappa_{zz,yx}/\kappa_{xy}$,
and (b) $\xi$ and $\kappa_{xy}$ versus temperature, based on the
fits (solid lines) shown in Fig. \ref{fig:T1Tinv}. Also shown are
the nematic susceptibilities measured by Raman and elastoresistance
measurements, reproduced from Refs. \protect\onlinecite{Ba122RamanPRL2013,Gallais2016}
and \protect\onlinecite{FisherScienceNematic2012}, respectively.
The solid lines are fits as described in the text.}
\end{figure}

The fits for $(T_{1}T)_{z}^{-1}$ and $(T_{1}T)_{y}^{-1}$ in the
absence of strain are shown as solid gray lines in Figs. \ref{fig:T1Tinv}(b)
and (d) for $\xi_{x}=\xi_{y}$. We find $\xi_{z}/\xi_{x}=0.88$, in
agreement with the fact that in the absence of strain the spins point
along the plane. Moreover, the temperature dependence of $\xi_{x}(T)$,
shown in Fig. \ref{fig:fitpars}(a), gives values consistent with
those measured by inelastic neutron scattering. Having fixed the unstrained
parameters, we perform fits in the presence of strain, shown by the
solid lines in Fig. \ref{fig:T1Tinv}(a) and (c). The only parameters
introduced in this case are the nematic order parameters $\varphi_{xy}=\varphi_{yx}$
and $\varphi_{zz}$. The good agreement between the fitted and the
experimental curves of both $(T_{1}T)_{z}^{-1}$ and $(T_{1}T)_{y}^{-1}$
over a wide temperature-strain regime demonstrates the suitability
of the phenomenological model employed in our analysis.

The temperature and strain behaviors of the nematic order parameters
$\varphi_{\alpha\beta}$ allows us to extract the temperature dependence
of the nematic susceptibilities $\chi_{\mathrm{nem}}^{(xy)}$ and
$\chi_{\mathrm{nem}}^{(zz)}$, as shown in Fig. \ref{fig:fitpars}(b).
It is clear that generally $\chi_{\mathrm{nem}}^{(zz)}>\chi_{\mathrm{nem}}^{(xy)}$,
particularly close to the magnetic transition. This quantitative analysis
corroborates the qualitative conclusion above, namely that nematic
order induces anisotropies in spin-space, and that the out-of-plane
spin fluctuations are more strongly enhanced by in-plane strain than
the in-plane spin fluctuations. It is interesting to compare $\chi_{\mathrm{nem}}^{(xy)}$
and $\chi_{\mathrm{nem}}^{(zz)}$ with the nematic susceptibility
extracted from elastoresistance \cite{FisherScienceNematic2012} and
from electronic Raman spectroscopy experiments \cite{Ba122RamanPRL2013}.
As shown in Fig. \ref{fig:fitpars}(b), the values are consistent,
and the NMR-extracted nematic susceptibilities also follow a Curie-Weiss
type of behavior \cite{Gallais2016}, with a Curie temperature $T_{0}=116$
K comparable to that extracted from the elastoresistance \cite{FisherScienceNematic2012}.
Note however that, in contrast to our NMR analysis, the other probes
for the nematic susceptibility are not sensitive to the ``polarization''
of the nematic susceptibility.

To the best of our knowledge, our results are the first to reveal
the internal spin structure of the nematic order parameter in iron-based
superconductors. This behavior is a clear manifestation of the entanglement
between spin, orbital, and lattice degrees of freedom in the normal
state of these compounds. Since superconductivity emerges from this
unique state, the rich interplay between these different degrees of
freedom revealed by our NMR analysis will certainly affect the properties
of the superconducting state.

The surprising anisotropic response of different nematic components
to in-plane strain reveals that the spin polarization can be controlled
by lattice distortions, similar to a piezomagnetic effect. In particular,
the result $\chi_{\mathrm{nem}}^{(zz)}>\chi_{\mathrm{nem}}^{(xy)}$
implies that for sufficiently large strain $\varepsilon^{*}$, the
dominant spin polarization will shift from in-plane to out-of-plane.
The value of $\varepsilon^{*}$ can be estimated from the condition
that the out-of-plane magnetic correlation length $\tilde{\xi}_{z}=\xi_{z}-\varepsilon\chi_{\mathrm{nem}}^{(zz)}$
becomes larger than the in-plane magnetic correlation length $\tilde{\xi}_{x}=\xi_{x}-\varepsilon\chi_{\mathrm{nem}}^{(xy)}$,
yielding $\varepsilon^{*}\approx0.4\%$ close to the magnetic transition
temperature. Such a strain value, which is just beyond the capability
of our specific piezo device, can  reasonably be achieved by
similar types of devices, however. More importantly, this analysis opens a
new avenue to control spin polarization in nematic materials without
using magnetic fields, but instead by using mechanical strain. Since
nematic order has been observed in other correlated materials such
as cuprates and ruthenates, it will be interesting to investigate
whether similar sizable effects are present in these systems as well.

More broadly, our work demonstrates that precision tunable strain
in combination with NMR provides a novel and important method to probe
spin and charge degrees of freedom. It provides an intriguing possibility
to tune the NMR spin relaxation rate by changing a voltage bias on
the piezoelectric stacks. The subtle coupling between the lattice
and spin polarizations exhibited by BaFe$_{2}$As$_{2}$ offers the
potential for controlling magnetic properties through lattice deformations
in next generation materials. Another potential application of our
technique is the use of nuclear quadrupolar resonance to image local
strains. The large response of the EFG to strain observed in this
study would translate into high spatial resolution in a linear strain
gradient, so that As NMR may be able to resolve microscopic features such as grain boundaries or defects.

\section{Acknowledgements}

We thank A. Dioguardi, S. Kivelson, and I. Fisher for enlightening
discussions, and P. Klavins, for assistance in the laboratory. Work
at UC Davis was supported by the NSF under Grant No.\ DMR-1506961.
RMF is supported by the U. S. Department of Energy, Office of Science,
Basic Energy Sciences, under award number DE-SC0012336. R. Sarkar was partially
supported by the DFG through SFB 1143 for the project C02. Work done
at Ames Lab (SLB, PCC, MT, RP, EIT) was supported by the U.S. Department
of Energy, Office of Basic Energy Science, Division of Materials Sciences
and Engineering. Ames Laboratory is operated for the U.S. Department
of Energy by Iowa State University under Contract No. DE-AC02-07CH11358.

\section{Methods}

\label{sec:methods} Crystals were grown in self-flux as described
in \cite{CanfieldBa122phasediagram2008} and cut along the $(110)_{{\rm T}}$
direction. Sample A had a mass of 2.52 mg and was mounted with the
field parallel to the $\mathbf{c}$ axis, and Sample B had a mass
0.91 mg and was mounted with the field perpendicular to the $\mathbf{c}$
axis (see Fig. \ref{fig:cartoon}). The crystals were secured with
heat-cured epoxy (UHU Plus 300 epoxy resin). Strain was applied along
the $(110)_{{\rm T}}$ direction using the CS100 cryogenic uniaxial
strain cell developed by Razorbill Instruments based on a design by
Hicks et. al. \cite{Hicks2014}, mounted in a modified probe operating
in a Quantum Design PPMS cryostat. The displacement, $x$ was measured
by monitoring the capacitance of using a precision capacitance bridge
with a resolution of 0.1nm. The strain was computed as $\epsilon=(x-x_{0})/L_{0}$,
where $L_{0}=2.052$ mm and $x_{0}=49.5$ $\mu$m for sample A and
$L_{0}=1.494$ mm and $x_{0}=51.58$ $\mu$m for sample B. For sample
B, positive (tensile) strain corresponds to $\mathbf{H}_{0}~||~\hat{b}$
and negative (compressive) strain corresponds to $\mathbf{H}_{0}~||~\hat{a}$.
Because the sample was mounted at room temperature, thermal contraction
creates positive strain even at zero piezo bias at low temperatures,
making a precise determination of $x_{0}$ difficult. For sample A
$x_{0}$ was determined by the minimum in $(T_{1}T)^{-1}$ versus
$x$, and for sample B $x_{0}$ was determined by the value $\nu_{bb}(x_{0})=|\nu_{cc}|/2=1.23$
MHz, where $\nu_{\alpha\alpha}$ is the quadrupolar splitting for
field along the $\alpha$ direction (see supplemental materials).
The maximum/minimum possible applied voltages to the piezoelectric
stacks limited the range of strains that could be applied to between
approximately $-0.002$ to $+0.003$ in the perpendicular case, and
$-0.0015$ to $+0.002$ for the parallel case. The spin-lattice relaxation
rate was measured using inversion recovery at the central transition
in fixed field, and the data were fit to the expression $M(t)=M_{0}\left[1-2f\left(\frac{9}{10}e^{-6t/T_{1}}+\frac{1}{10}e^{-t/T_{1}}\right)\right]$.
The data were well-fit to a single $T_{1}$ value.

\bibliography{NMRstrainBibliography}

\begin{thebibliography}{32}%
\makeatletter
\providecommand \@ifxundefined [1]{%
 \@ifx{#1\undefined}
}%
\providecommand \@ifnum [1]{%
 \ifnum #1\expandafter \@firstoftwo
 \else \expandafter \@secondoftwo
 \fi
}%
\providecommand \@ifx [1]{%
 \ifx #1\expandafter \@firstoftwo
 \else \expandafter \@secondoftwo
 \fi
}%
\providecommand \natexlab [1]{#1}%
\providecommand \enquote  [1]{``#1''}%
\providecommand \bibnamefont  [1]{#1}%
\providecommand \bibfnamefont [1]{#1}%
\providecommand \citenamefont [1]{#1}%
\providecommand \href@noop [0]{\@secondoftwo}%
\providecommand \href [0]{\begingroup \@sanitize@url \@href}%
\providecommand \@href[1]{\@@startlink{#1}\@@href}%
\providecommand \@@href[1]{\endgroup#1\@@endlink}%
\providecommand \@sanitize@url [0]{\catcode `\\12\catcode `\$12\catcode
  `\&12\catcode `\#12\catcode `\^12\catcode `\_12\catcode `\%12\relax}%
\providecommand \@@startlink[1]{}%
\providecommand \@@endlink[0]{}%
\providecommand \url  [0]{\begingroup\@sanitize@url \@url }%
\providecommand \@url [1]{\endgroup\@href {#1}{\urlprefix }}%
\providecommand \urlprefix  [0]{URL }%
\providecommand \Eprint [0]{\href }%
\providecommand \doibase [0]{http://dx.doi.org/}%
\providecommand \selectlanguage [0]{\@gobble}%
\providecommand \bibinfo  [0]{\@secondoftwo}%
\providecommand \bibfield  [0]{\@secondoftwo}%
\providecommand \translation [1]{[#1]}%
\providecommand \BibitemOpen [0]{}%
\providecommand \bibitemStop [0]{}%
\providecommand \bibitemNoStop [0]{.\EOS\space}%
\providecommand \EOS [0]{\spacefactor3000\relax}%
\providecommand \BibitemShut  [1]{\csname bibitem#1\endcsname}%
\let\auto@bib@innerbib\@empty
\bibitem [{\citenamefont {Canfield}\ and\ \citenamefont
  {Bud'ko}(2010)}]{doping122review}%
  \BibitemOpen
  \bibfield  {author} {\bibinfo {author} {\bibfnamefont {Paul~C.}\ \bibnamefont
  {Canfield}}\ and\ \bibinfo {author} {\bibfnamefont {Sergey~L.}\ \bibnamefont
  {Bud'ko}},\ }\bibfield  {title} {\enquote {\bibinfo {title} {{FeAs}-based
  superconductivity: A case study of the effects of transition metal doping on
  {BaFe$_2$As$_2$}},}\ }\href {\doibase
  10.1146/annurev-conmatphys-070909-104041} {\bibfield  {journal} {\bibinfo
  {journal} {Annu. Rev. Condens. Matter Phys.}\ }\textbf {\bibinfo {volume}
  {1}},\ \bibinfo {pages} {27--50} (\bibinfo {year} {2010})}\BibitemShut
  {NoStop}%
\bibitem [{\citenamefont {Fernandes}\ \emph {et~al.}(2012)\citenamefont
  {Fernandes}, \citenamefont {Chubukov}, \citenamefont {Knolle}, \citenamefont
  {Eremin},\ and\ \citenamefont {Schmalian}}]{FernandesNematicPnictides}%
  \BibitemOpen
  \bibfield  {author} {\bibinfo {author} {\bibfnamefont {R.~M.}\ \bibnamefont
  {Fernandes}}, \bibinfo {author} {\bibfnamefont {A.~V.}\ \bibnamefont
  {Chubukov}}, \bibinfo {author} {\bibfnamefont {J.}~\bibnamefont {Knolle}},
  \bibinfo {author} {\bibfnamefont {I.}~\bibnamefont {Eremin}}, \ and\ \bibinfo
  {author} {\bibfnamefont {J.}~\bibnamefont {Schmalian}},\ }\bibfield  {title}
  {\enquote {\bibinfo {title} {Preemptive nematic order, pseudogap, and orbital
  order in the iron pnictides},}\ }\href {\doibase 10.1103/PhysRevB.85.024534}
  {\bibfield  {journal} {\bibinfo  {journal} {Phys. Rev. B}\ }\textbf {\bibinfo
  {volume} {85}},\ \bibinfo {pages} {024534} (\bibinfo {year}
  {2012})}\BibitemShut {NoStop}%
\bibitem [{\citenamefont {Fernandes}\ and\ \citenamefont
  {Schmalian}(2012)}]{Fernandes2012}%
  \BibitemOpen
  \bibfield  {author} {\bibinfo {author} {\bibfnamefont {Rafael~M}\
  \bibnamefont {Fernandes}}\ and\ \bibinfo {author} {\bibfnamefont {JÃ¶rg}\
  \bibnamefont {Schmalian}},\ }\bibfield  {title} {\enquote {\bibinfo {title}
  {Manifestations of nematic degrees of freedom in the magnetic, elastic, and
  superconducting properties of the iron pnictides},}\ }\href {\doibase
  10.1088/0953-2048/25/8/084005} {\bibfield  {journal} {\bibinfo  {journal}
  {Supercond. Sci. Technol.}\ }\textbf {\bibinfo {volume} {25}},\ \bibinfo
  {pages} {084005} (\bibinfo {year} {2012})}\BibitemShut {NoStop}%
\bibitem [{\citenamefont {Fernandes}\ \emph {et~al.}(2013)\citenamefont
  {Fernandes}, \citenamefont {B\"ohmer}, \citenamefont {Meingast},\ and\
  \citenamefont {Schmalian}}]{FernandesPRLnematicT1}%
  \BibitemOpen
  \bibfield  {author} {\bibinfo {author} {\bibfnamefont {Rafael~M.}\
  \bibnamefont {Fernandes}}, \bibinfo {author} {\bibfnamefont {Anna~E.}\
  \bibnamefont {B\"ohmer}}, \bibinfo {author} {\bibfnamefont {Christoph}\
  \bibnamefont {Meingast}}, \ and\ \bibinfo {author} {\bibfnamefont {J\"org}\
  \bibnamefont {Schmalian}},\ }\bibfield  {title} {\enquote {\bibinfo {title}
  {Scaling between magnetic and lattice fluctuations in iron pnictide
  superconductors},}\ }\href {\doibase 10.1103/PhysRevLett.111.137001}
  {\bibfield  {journal} {\bibinfo  {journal} {Phys. Rev. Lett.}\ }\textbf
  {\bibinfo {volume} {111}},\ \bibinfo {pages} {137001} (\bibinfo {year}
  {2013})}\BibitemShut {NoStop}%
\bibitem [{\citenamefont {Blomberg}\ \emph {et~al.}(2012)\citenamefont
  {Blomberg}, \citenamefont {Kreyssig}, \citenamefont {Tanatar}, \citenamefont
  {Fernandes}, \citenamefont {Kim}, \citenamefont {Thaler}, \citenamefont
  {Schmalian}, \citenamefont {Bud'ko}, \citenamefont {Canfield}, \citenamefont
  {Goldman},\ and\ \citenamefont {Prozorov}}]{TanatarTensileStressPRB}%
  \BibitemOpen
  \bibfield  {author} {\bibinfo {author} {\bibfnamefont {E.~C.}\ \bibnamefont
  {Blomberg}}, \bibinfo {author} {\bibfnamefont {A.}~\bibnamefont {Kreyssig}},
  \bibinfo {author} {\bibfnamefont {M.~A.}\ \bibnamefont {Tanatar}}, \bibinfo
  {author} {\bibfnamefont {R.~M.}\ \bibnamefont {Fernandes}}, \bibinfo {author}
  {\bibfnamefont {M.~G.}\ \bibnamefont {Kim}}, \bibinfo {author} {\bibfnamefont
  {A.}~\bibnamefont {Thaler}}, \bibinfo {author} {\bibfnamefont
  {J.}~\bibnamefont {Schmalian}}, \bibinfo {author} {\bibfnamefont {S.~L.}\
  \bibnamefont {Bud'ko}}, \bibinfo {author} {\bibfnamefont {P.~C.}\
  \bibnamefont {Canfield}}, \bibinfo {author} {\bibfnamefont {A.~I.}\
  \bibnamefont {Goldman}}, \ and\ \bibinfo {author} {\bibfnamefont
  {R.}~\bibnamefont {Prozorov}},\ }\bibfield  {title} {\enquote {\bibinfo
  {title} {Effect of tensile stress on the in-plane resistivity anisotropy in
  {BaFe${}_{2}$As${}_{2}$}},}\ }\href {\doibase 10.1103/PhysRevB.85.144509}
  {\bibfield  {journal} {\bibinfo  {journal} {Phys. Rev. B}\ }\textbf {\bibinfo
  {volume} {85}},\ \bibinfo {pages} {144509} (\bibinfo {year}
  {2012})}\BibitemShut {NoStop}%
\bibitem [{\citenamefont {Lu}\ \emph {et~al.}(2014)\citenamefont {Lu},
  \citenamefont {Park}, \citenamefont {Zhang}, \citenamefont {Luo},
  \citenamefont {Nevidomskyy}, \citenamefont {Si},\ and\ \citenamefont
  {Dai}}]{StrainedPnictidesNS2014science}%
  \BibitemOpen
  \bibfield  {author} {\bibinfo {author} {\bibfnamefont {Xingye}\ \bibnamefont
  {Lu}}, \bibinfo {author} {\bibfnamefont {J.~T.}\ \bibnamefont {Park}},
  \bibinfo {author} {\bibfnamefont {Rui}\ \bibnamefont {Zhang}}, \bibinfo
  {author} {\bibfnamefont {Huiqian}\ \bibnamefont {Luo}}, \bibinfo {author}
  {\bibfnamefont {Andriy~H.}\ \bibnamefont {Nevidomskyy}}, \bibinfo {author}
  {\bibfnamefont {Qimiao}\ \bibnamefont {Si}}, \ and\ \bibinfo {author}
  {\bibfnamefont {Pengcheng}\ \bibnamefont {Dai}},\ }\bibfield  {title}
  {\enquote {\bibinfo {title} {Nematic spin correlations in the tetragonal
  state of uniaxial-strained {BaFe$_{2-x}$Ni$_x$As$_2$}},}\ }\href {\doibase
  10.1126/science.1251853} {\bibfield  {journal} {\bibinfo  {journal}
  {Science}\ }\textbf {\bibinfo {volume} {345}},\ \bibinfo {pages} {657 -- 660}
  (\bibinfo {year} {2014})}\BibitemShut {NoStop}%
\bibitem [{\citenamefont {Man}\ \emph {et~al.}(2015)\citenamefont {Man},
  \citenamefont {Lu}, \citenamefont {Chen}, \citenamefont {Zhang},
  \citenamefont {Zhang}, \citenamefont {Luo}, \citenamefont {Kulda},
  \citenamefont {Ivanov}, \citenamefont {Keller}, \citenamefont {Morosan},
  \citenamefont {Si},\ and\ \citenamefont {Dai}}]{StrainBa122neutronsPRB2015}%
  \BibitemOpen
  \bibfield  {author} {\bibinfo {author} {\bibfnamefont {Haoran}\ \bibnamefont
  {Man}}, \bibinfo {author} {\bibfnamefont {Xingye}\ \bibnamefont {Lu}},
  \bibinfo {author} {\bibfnamefont {Justin~S.}\ \bibnamefont {Chen}}, \bibinfo
  {author} {\bibfnamefont {Rui}\ \bibnamefont {Zhang}}, \bibinfo {author}
  {\bibfnamefont {Wenliang}\ \bibnamefont {Zhang}}, \bibinfo {author}
  {\bibfnamefont {Huiqian}\ \bibnamefont {Luo}}, \bibinfo {author}
  {\bibfnamefont {J.}~\bibnamefont {Kulda}}, \bibinfo {author} {\bibfnamefont
  {A.}~\bibnamefont {Ivanov}}, \bibinfo {author} {\bibfnamefont
  {T.}~\bibnamefont {Keller}}, \bibinfo {author} {\bibfnamefont {Emilia}\
  \bibnamefont {Morosan}}, \bibinfo {author} {\bibfnamefont {Qimiao}\
  \bibnamefont {Si}}, \ and\ \bibinfo {author} {\bibfnamefont {Pengcheng}\
  \bibnamefont {Dai}},\ }\bibfield  {title} {\enquote {\bibinfo {title}
  {Electronic nematic correlations in the stress-free tetragonal state of
  {${\mathrm{BaFe}}_{2\ensuremath{-}x}{\mathrm{Ni}}_{x}{\mathrm{As}}_{2}$}},}\
  }\href {\doibase 10.1103/PhysRevB.92.134521} {\bibfield  {journal} {\bibinfo
  {journal} {Phys. Rev. B}\ }\textbf {\bibinfo {volume} {92}},\ \bibinfo
  {pages} {134521} (\bibinfo {year} {2015})}\BibitemShut {NoStop}%
\bibitem [{\citenamefont {Fradkin}\ \emph {et~al.}(2010)\citenamefont
  {Fradkin}, \citenamefont {Kivelson}, \citenamefont {Lawler}, \citenamefont
  {Eisenstein},\ and\ \citenamefont {Mackenzie}}]{FradkinNematicReview}%
  \BibitemOpen
  \bibfield  {author} {\bibinfo {author} {\bibfnamefont {Eduardo}\ \bibnamefont
  {Fradkin}}, \bibinfo {author} {\bibfnamefont {Steven~A.}\ \bibnamefont
  {Kivelson}}, \bibinfo {author} {\bibfnamefont {Michael~J.}\ \bibnamefont
  {Lawler}}, \bibinfo {author} {\bibfnamefont {James~P.}\ \bibnamefont
  {Eisenstein}}, \ and\ \bibinfo {author} {\bibfnamefont {Andrew~P.}\
  \bibnamefont {Mackenzie}},\ }\bibfield  {title} {\enquote {\bibinfo {title}
  {Nematic fermi fluids in condensed matter physics},}\ }\href {\doibase
  10.1146/annurev-conmatphys-070909-103925} {\bibfield  {journal} {\bibinfo
  {journal} {Annu. Rev. Condens. Matter Phys.}\ }\textbf {\bibinfo {volume}
  {1}},\ \bibinfo {pages} {153--178} (\bibinfo {year} {2010})}\BibitemShut
  {NoStop}%
\bibitem [{\citenamefont {Tanatar}\ \emph {et~al.}(2010)\citenamefont
  {Tanatar}, \citenamefont {Blomberg}, \citenamefont {Kreyssig}, \citenamefont
  {Kim}, \citenamefont {Ni}, \citenamefont {Thaler}, \citenamefont {Bud'ko},
  \citenamefont {Canfield}, \citenamefont {Goldman}, \citenamefont {Mazin},\
  and\ \citenamefont {Prozorov}}]{TanatarDetwin}%
  \BibitemOpen
  \bibfield  {author} {\bibinfo {author} {\bibfnamefont {M.~A.}\ \bibnamefont
  {Tanatar}}, \bibinfo {author} {\bibfnamefont {E.~C.}\ \bibnamefont
  {Blomberg}}, \bibinfo {author} {\bibfnamefont {A.}~\bibnamefont {Kreyssig}},
  \bibinfo {author} {\bibfnamefont {M.~G.}\ \bibnamefont {Kim}}, \bibinfo
  {author} {\bibfnamefont {N.}~\bibnamefont {Ni}}, \bibinfo {author}
  {\bibfnamefont {A.}~\bibnamefont {Thaler}}, \bibinfo {author} {\bibfnamefont
  {S.~L.}\ \bibnamefont {Bud'ko}}, \bibinfo {author} {\bibfnamefont {P.~C.}\
  \bibnamefont {Canfield}}, \bibinfo {author} {\bibfnamefont {A.~I.}\
  \bibnamefont {Goldman}}, \bibinfo {author} {\bibfnamefont {I.~I.}\
  \bibnamefont {Mazin}}, \ and\ \bibinfo {author} {\bibfnamefont
  {R.}~\bibnamefont {Prozorov}},\ }\bibfield  {title} {\enquote {\bibinfo
  {title} {Uniaxial-strain mechanical detwinning of {CaFe$_2$As$_2$} and
  {BaFe$_2$As$_2$} crystals: Optical and transport study},}\ }\href {\doibase
  10.1103/PhysRevB.81.184508} {\bibfield  {journal} {\bibinfo  {journal} {Phys.
  Rev. B}\ }\textbf {\bibinfo {volume} {81}},\ \bibinfo {pages} {184508}
  (\bibinfo {year} {2010})}\BibitemShut {NoStop}%
\bibitem [{\citenamefont {Chu}\ \emph {et~al.}(2010)\citenamefont {Chu},
  \citenamefont {Analytis}, \citenamefont {De~Greve}, \citenamefont {McMahon},
  \citenamefont {Islam}, \citenamefont {Yamamoto},\ and\ \citenamefont
  {Fisher}}]{IronArsenideDetwinnedFisherScience2010}%
  \BibitemOpen
  \bibfield  {author} {\bibinfo {author} {\bibfnamefont {Jiun-Haw}\
  \bibnamefont {Chu}}, \bibinfo {author} {\bibfnamefont {James~G.}\
  \bibnamefont {Analytis}}, \bibinfo {author} {\bibfnamefont {Kristiaan}\
  \bibnamefont {De~Greve}}, \bibinfo {author} {\bibfnamefont {Peter~L.}\
  \bibnamefont {McMahon}}, \bibinfo {author} {\bibfnamefont {Zahirul}\
  \bibnamefont {Islam}}, \bibinfo {author} {\bibfnamefont {Yoshihisa}\
  \bibnamefont {Yamamoto}}, \ and\ \bibinfo {author} {\bibfnamefont {Ian~R.}\
  \bibnamefont {Fisher}},\ }\bibfield  {title} {\enquote {\bibinfo {title}
  {In-plane resistivity anisotropy in an underdoped iron arsenide
  superconductor},}\ }\href {\doibase 10.1126/science.1190482} {\bibfield
  {journal} {\bibinfo  {journal} {Science}\ }\textbf {\bibinfo {volume}
  {329}},\ \bibinfo {pages} {824--826} (\bibinfo {year} {2010})}\BibitemShut
  {NoStop}%
\bibitem [{\citenamefont {Fernandes}\ \emph {et~al.}(2014)\citenamefont
  {Fernandes}, \citenamefont {Chubukov},\ and\ \citenamefont
  {Schmalian}}]{FernandesSchmalianNatPhys2014}%
  \BibitemOpen
  \bibfield  {author} {\bibinfo {author} {\bibfnamefont {R.~M.}\ \bibnamefont
  {Fernandes}}, \bibinfo {author} {\bibfnamefont {A.~V.}\ \bibnamefont
  {Chubukov}}, \ and\ \bibinfo {author} {\bibfnamefont {J.}~\bibnamefont
  {Schmalian}},\ }\bibfield  {title} {\enquote {\bibinfo {title} {What drives
  nematic order in iron-based superconductors?}}\ }\href
  {http://dx.doi.org/10.1038/nphys2877} {\bibfield  {journal} {\bibinfo
  {journal} {Nat. Phys.}\ }\textbf {\bibinfo {volume} {10}},\ \bibinfo {pages}
  {97--104} (\bibinfo {year} {2014})}\BibitemShut {NoStop}%
\bibitem [{\citenamefont {Dai}(2015)}]{DaiRMP2015}%
  \BibitemOpen
  \bibfield  {author} {\bibinfo {author} {\bibfnamefont {Pengcheng}\
  \bibnamefont {Dai}},\ }\bibfield  {title} {\enquote {\bibinfo {title}
  {Antiferromagnetic order and spin dynamics in iron-based superconductors},}\
  }\href {\doibase 10.1103/RevModPhys.87.855} {\bibfield  {journal} {\bibinfo
  {journal} {Rev. Mod. Phys.}\ }\textbf {\bibinfo {volume} {87}},\ \bibinfo
  {pages} {855--896} (\bibinfo {year} {2015})}\BibitemShut {NoStop}%
\bibitem [{\citenamefont {Song}\ \emph {et~al.}(2015)\citenamefont {Song},
  \citenamefont {Lu}, \citenamefont {Abernathy}, \citenamefont {Tam},
  \citenamefont {Niedziela}, \citenamefont {Tian}, \citenamefont {Luo},
  \citenamefont {Si},\ and\ \citenamefont {Dai}}]{PengchangPRB2015}%
  \BibitemOpen
  \bibfield  {author} {\bibinfo {author} {\bibfnamefont {Yu}~\bibnamefont
  {Song}}, \bibinfo {author} {\bibfnamefont {Xingye}\ \bibnamefont {Lu}},
  \bibinfo {author} {\bibfnamefont {D.~L.}\ \bibnamefont {Abernathy}}, \bibinfo
  {author} {\bibfnamefont {David~W.}\ \bibnamefont {Tam}}, \bibinfo {author}
  {\bibfnamefont {J.~L.}\ \bibnamefont {Niedziela}}, \bibinfo {author}
  {\bibfnamefont {Wei}\ \bibnamefont {Tian}}, \bibinfo {author} {\bibfnamefont
  {Huiqian}\ \bibnamefont {Luo}}, \bibinfo {author} {\bibfnamefont {Qimiao}\
  \bibnamefont {Si}}, \ and\ \bibinfo {author} {\bibfnamefont {Pengcheng}\
  \bibnamefont {Dai}},\ }\bibfield  {title} {\enquote {\bibinfo {title} {Energy
  dependence of the spin excitation anisotropy in uniaxial-strained
  {${\mathrm{BaFe}}_{1.9}{\mathrm{Ni}}_{0.1}{\mathrm{As}}_{2}$}},}\ }\href
  {\doibase 10.1103/PhysRevB.92.180504} {\bibfield  {journal} {\bibinfo
  {journal} {Phys. Rev. B}\ }\textbf {\bibinfo {volume} {92}},\ \bibinfo
  {pages} {180504} (\bibinfo {year} {2015})}\BibitemShut {NoStop}%
\bibitem [{\citenamefont {Dioguardi}\ \emph {et~al.}(2015)\citenamefont
  {Dioguardi}, \citenamefont {Lawson}, \citenamefont {Bush}, \citenamefont
  {Crocker}, \citenamefont {Shirer}, \citenamefont {Nisson}, \citenamefont
  {Kissikov}, \citenamefont {Ran}, \citenamefont {Bud'ko}, \citenamefont
  {Canfield}, \citenamefont {Yuan}, \citenamefont {Kuhns}, \citenamefont
  {Reyes}, \citenamefont {Grafe},\ and\ \citenamefont
  {Curro}}]{DioguardiNematicGlass2015}%
  \BibitemOpen
  \bibfield  {author} {\bibinfo {author} {\bibfnamefont {A.~P.}\ \bibnamefont
  {Dioguardi}}, \bibinfo {author} {\bibfnamefont {M.~M.}\ \bibnamefont
  {Lawson}}, \bibinfo {author} {\bibfnamefont {B.~T.}\ \bibnamefont {Bush}},
  \bibinfo {author} {\bibfnamefont {J.}~\bibnamefont {Crocker}}, \bibinfo
  {author} {\bibfnamefont {K.~R.}\ \bibnamefont {Shirer}}, \bibinfo {author}
  {\bibfnamefont {D.~M.}\ \bibnamefont {Nisson}}, \bibinfo {author}
  {\bibfnamefont {T.}~\bibnamefont {Kissikov}}, \bibinfo {author}
  {\bibfnamefont {S.}~\bibnamefont {Ran}}, \bibinfo {author} {\bibfnamefont
  {S.~L.}\ \bibnamefont {Bud'ko}}, \bibinfo {author} {\bibfnamefont {P.~C.}\
  \bibnamefont {Canfield}}, \bibinfo {author} {\bibfnamefont {S.}~\bibnamefont
  {Yuan}}, \bibinfo {author} {\bibfnamefont {P.~L.}\ \bibnamefont {Kuhns}},
  \bibinfo {author} {\bibfnamefont {A.~P.}\ \bibnamefont {Reyes}}, \bibinfo
  {author} {\bibfnamefont {H.-J.}\ \bibnamefont {Grafe}}, \ and\ \bibinfo
  {author} {\bibfnamefont {N.~J.}\ \bibnamefont {Curro}},\ }\bibfield  {title}
  {\enquote {\bibinfo {title} {{NMR} evidence for inhomogeneous glassy behavior
  driven by nematic fluctuations in iron arsenide superconductors},}\ }\href
  {\doibase 10.1103/PhysRevB.92.165116} {\bibfield  {journal} {\bibinfo
  {journal} {Phys. Rev. B}\ }\textbf {\bibinfo {volume} {92}},\ \bibinfo
  {pages} {165116} (\bibinfo {year} {2015})}\BibitemShut {NoStop}%
\bibitem [{\citenamefont {Kissikov}\ \emph {et~al.}(2016)\citenamefont
  {Kissikov}, \citenamefont {Dioguardi}, \citenamefont {Timmons}, \citenamefont
  {Tanatar}, \citenamefont {Prozorov}, \citenamefont {Bud'ko}, \citenamefont
  {Canfield}, \citenamefont {Fernandes},\ and\ \citenamefont
  {Curro}}]{NMRnematicStrainBa122PRB2016}%
  \BibitemOpen
  \bibfield  {author} {\bibinfo {author} {\bibfnamefont {T.}~\bibnamefont
  {Kissikov}}, \bibinfo {author} {\bibfnamefont {A.~P.}\ \bibnamefont
  {Dioguardi}}, \bibinfo {author} {\bibfnamefont {E.~I.}\ \bibnamefont
  {Timmons}}, \bibinfo {author} {\bibfnamefont {M.~A.}\ \bibnamefont
  {Tanatar}}, \bibinfo {author} {\bibfnamefont {R.}~\bibnamefont {Prozorov}},
  \bibinfo {author} {\bibfnamefont {S.~L.}\ \bibnamefont {Bud'ko}}, \bibinfo
  {author} {\bibfnamefont {P.~C.}\ \bibnamefont {Canfield}}, \bibinfo {author}
  {\bibfnamefont {R.~M.}\ \bibnamefont {Fernandes}}, \ and\ \bibinfo {author}
  {\bibfnamefont {N.~J.}\ \bibnamefont {Curro}},\ }\bibfield  {title} {\enquote
  {\bibinfo {title} {{NMR} study of nematic spin fluctuations in a detwinned
  single crystal of underdoped {Ba(Fe$_{1-x}$Co$_x$)$_2$As$_2$}},}\ }\href
  {\doibase 10.1103/PhysRevB.94.165123} {\bibfield  {journal} {\bibinfo
  {journal} {Phys. Rev. B}\ }\textbf {\bibinfo {volume} {94}},\ \bibinfo
  {pages} {165123} (\bibinfo {year} {2016})}\BibitemShut {NoStop}%
\bibitem [{\citenamefont {Ning}\ \emph {et~al.}(2014)\citenamefont {Ning},
  \citenamefont {Fu}, \citenamefont {Torchetti}, \citenamefont {Imai},
  \citenamefont {Sefat}, \citenamefont {Cheng}, \citenamefont {Shen},\ and\
  \citenamefont {Wen}}]{PhysRevB.89.214511}%
  \BibitemOpen
  \bibfield  {author} {\bibinfo {author} {\bibfnamefont {F.~L.}\ \bibnamefont
  {Ning}}, \bibinfo {author} {\bibfnamefont {M.}~\bibnamefont {Fu}}, \bibinfo
  {author} {\bibfnamefont {D.~A.}\ \bibnamefont {Torchetti}}, \bibinfo {author}
  {\bibfnamefont {T.}~\bibnamefont {Imai}}, \bibinfo {author} {\bibfnamefont
  {A.~S.}\ \bibnamefont {Sefat}}, \bibinfo {author} {\bibfnamefont
  {P.}~\bibnamefont {Cheng}}, \bibinfo {author} {\bibfnamefont
  {B.}~\bibnamefont {Shen}}, \ and\ \bibinfo {author} {\bibfnamefont {H.-H.}\
  \bibnamefont {Wen}},\ }\bibfield  {title} {\enquote {\bibinfo {title}
  {Critical behavior of the spin density wave transition in underdoped
  {Ba(Fe$_{1-x}$Co$_x$)$_2$As$_2$} ($x\ensuremath{\le}0.05$): $^{75}${As} {NMR}
  investigation},}\ }\href {\doibase 10.1103/PhysRevB.89.214511} {\bibfield
  {journal} {\bibinfo  {journal} {Phys. Rev. B}\ }\textbf {\bibinfo {volume}
  {89}},\ \bibinfo {pages} {214511} (\bibinfo {year} {2014})}\BibitemShut
  {NoStop}%
\bibitem [{\citenamefont {Smerald}\ and\ \citenamefont
  {Shannon}(2011)}]{T1formfactorsArsenides}%
  \BibitemOpen
  \bibfield  {author} {\bibinfo {author} {\bibfnamefont {Andrew}\ \bibnamefont
  {Smerald}}\ and\ \bibinfo {author} {\bibfnamefont {Nic}\ \bibnamefont
  {Shannon}},\ }\bibfield  {title} {\enquote {\bibinfo {title} {Angle-resolved
  {NMR}: Quantitative theory of ${}^{75}${As} ${T}_{1}$ relaxation rate in
  {BaFe${}_{2}$As${}_{2}$}},}\ }\href {\doibase 10.1103/PhysRevB.84.184437}
  {\bibfield  {journal} {\bibinfo  {journal} {Phys. Rev. B}\ }\textbf {\bibinfo
  {volume} {84}},\ \bibinfo {pages} {184437} (\bibinfo {year}
  {2011})}\BibitemShut {NoStop}%
\bibitem [{\citenamefont {Hicks}\ \emph
  {et~al.}(2014{\natexlab{a}})\citenamefont {Hicks}, \citenamefont {Barber},
  \citenamefont {Edkins}, \citenamefont {Brodsky},\ and\ \citenamefont
  {Mackenzie}}]{Hicks2014}%
  \BibitemOpen
  \bibfield  {author} {\bibinfo {author} {\bibfnamefont {Clifford~W.}\
  \bibnamefont {Hicks}}, \bibinfo {author} {\bibfnamefont {Mark~E.}\
  \bibnamefont {Barber}}, \bibinfo {author} {\bibfnamefont {Stephen~D.}\
  \bibnamefont {Edkins}}, \bibinfo {author} {\bibfnamefont {Daniel~O.}\
  \bibnamefont {Brodsky}}, \ and\ \bibinfo {author} {\bibfnamefont {Andrew~P.}\
  \bibnamefont {Mackenzie}},\ }\bibfield  {title} {\enquote {\bibinfo {title}
  {Piezoelectric-based apparatus for strain tuning},}\ }\href {\doibase
  10.1063/1.4881611} {\bibfield  {journal} {\bibinfo  {journal} {Rev. Sci.
  Instrum.}\ }\textbf {\bibinfo {volume} {85}},\ \bibinfo {pages} {065003}
  (\bibinfo {year} {2014}{\natexlab{a}})}\BibitemShut {NoStop}%
\bibitem [{\citenamefont {Hicks}\ \emph
  {et~al.}(2014{\natexlab{b}})\citenamefont {Hicks}, \citenamefont {Brodsky},
  \citenamefont {Yelland}, \citenamefont {Gibbs}, \citenamefont {Bruin},
  \citenamefont {Barber}, \citenamefont {Edkins}, \citenamefont {Nishimura},
  \citenamefont {Yonezawa}, \citenamefont {Maeno},\ and\ \citenamefont
  {Mackenzie}}]{Sr2RuO4strainScience2014}%
  \BibitemOpen
  \bibfield  {author} {\bibinfo {author} {\bibfnamefont {C.~W.}\ \bibnamefont
  {Hicks}}, \bibinfo {author} {\bibfnamefont {D.~O.}\ \bibnamefont {Brodsky}},
  \bibinfo {author} {\bibfnamefont {E.~A.}\ \bibnamefont {Yelland}}, \bibinfo
  {author} {\bibfnamefont {A.~S.}\ \bibnamefont {Gibbs}}, \bibinfo {author}
  {\bibfnamefont {J.~A.~N.}\ \bibnamefont {Bruin}}, \bibinfo {author}
  {\bibfnamefont {M.~E.}\ \bibnamefont {Barber}}, \bibinfo {author}
  {\bibfnamefont {S.~D.}\ \bibnamefont {Edkins}}, \bibinfo {author}
  {\bibfnamefont {K.}~\bibnamefont {Nishimura}}, \bibinfo {author}
  {\bibfnamefont {S.}~\bibnamefont {Yonezawa}}, \bibinfo {author}
  {\bibfnamefont {Y.}~\bibnamefont {Maeno}}, \ and\ \bibinfo {author}
  {\bibfnamefont {A.~P.}\ \bibnamefont {Mackenzie}},\ }\bibfield  {title}
  {\enquote {\bibinfo {title} {Strong increase of {T$_c$} of {Sr$_2$RuO$_4$}
  under both tensile and compressive strain},}\ }\href {\doibase
  10.1126/science.1248292} {\bibfield  {journal} {\bibinfo  {journal}
  {Science}\ }\textbf {\bibinfo {volume} {344}},\ \bibinfo {pages} {283--285}
  (\bibinfo {year} {2014}{\natexlab{b}})}\BibitemShut {NoStop}%
\bibitem [{\citenamefont {Kitagawa}\ \emph {et~al.}(2008)\citenamefont
  {Kitagawa}, \citenamefont {Katayama}, \citenamefont {Ohgushi}, \citenamefont
  {Yoshida},\ and\ \citenamefont {Takigawa}}]{takigawa2008}%
  \BibitemOpen
  \bibfield  {author} {\bibinfo {author} {\bibfnamefont {Kentaro}\ \bibnamefont
  {Kitagawa}}, \bibinfo {author} {\bibfnamefont {Naoyuki}\ \bibnamefont
  {Katayama}}, \bibinfo {author} {\bibfnamefont {Kenya}\ \bibnamefont
  {Ohgushi}}, \bibinfo {author} {\bibfnamefont {Makoto}\ \bibnamefont
  {Yoshida}}, \ and\ \bibinfo {author} {\bibfnamefont {Masashi}\ \bibnamefont
  {Takigawa}},\ }\bibfield  {title} {\enquote {\bibinfo {title} {Commensurate
  itinerant antiferromagnetism in {BaFe$_{2}$As$_{2}$: $^{75}$As-NMR} studies
  on a self-flux grown single crystal},}\ }\href {\doibase
  10.1143/JPSJ.77.114709} {\bibfield  {journal} {\bibinfo  {journal} {J. Phys.
  Soc. Jpn.}\ }\textbf {\bibinfo {volume} {77}},\ \bibinfo {pages} {114709}
  (\bibinfo {year} {2008})}\BibitemShut {NoStop}%
\bibitem [{\citenamefont {Gallais}\ \emph {et~al.}(2013)\citenamefont
  {Gallais}, \citenamefont {Fernandes}, \citenamefont {Paul}, \citenamefont
  {Chauvi{\`{e}}re}, \citenamefont {Yang}, \citenamefont {M{\'{e}}asson},
  \citenamefont {Cazayous}, \citenamefont {Sacuto}, \citenamefont {Colson},\
  and\ \citenamefont {Forget}}]{Ba122RamanPRL2013}%
  \BibitemOpen
  \bibfield  {author} {\bibinfo {author} {\bibfnamefont {Y.}~\bibnamefont
  {Gallais}}, \bibinfo {author} {\bibfnamefont {R.~M.}\ \bibnamefont
  {Fernandes}}, \bibinfo {author} {\bibfnamefont {I.}~\bibnamefont {Paul}},
  \bibinfo {author} {\bibfnamefont {L.}~\bibnamefont {Chauvi{\`{e}}re}},
  \bibinfo {author} {\bibfnamefont {Y.-X.}\ \bibnamefont {Yang}}, \bibinfo
  {author} {\bibfnamefont {M.-A.}\ \bibnamefont {M{\'{e}}asson}}, \bibinfo
  {author} {\bibfnamefont {M.}~\bibnamefont {Cazayous}}, \bibinfo {author}
  {\bibfnamefont {A.}~\bibnamefont {Sacuto}}, \bibinfo {author} {\bibfnamefont
  {D.}~\bibnamefont {Colson}}, \ and\ \bibinfo {author} {\bibfnamefont
  {A.}~\bibnamefont {Forget}},\ }\bibfield  {title} {\enquote {\bibinfo {title}
  {Observation of incipient charge nematicity in
  {Ba(Fe$_{1-x}$Co$_x$)$_2$As$_2$}},}\ }\href {\doibase
  10.1103/PhysRevLett.111.267001} {\bibfield  {journal} {\bibinfo  {journal}
  {Phys. Rev. Lett.}\ }\textbf {\bibinfo {volume} {111}},\ \bibinfo {pages}
  {267001} (\bibinfo {year} {2013})}\BibitemShut {NoStop}%
\bibitem [{\citenamefont {Gallais}\ and\ \citenamefont
  {Paul}(2016)}]{Gallais2016}%
  \BibitemOpen
  \bibfield  {author} {\bibinfo {author} {\bibfnamefont {Yann}\ \bibnamefont
  {Gallais}}\ and\ \bibinfo {author} {\bibfnamefont {Indranil}\ \bibnamefont
  {Paul}},\ }\bibfield  {title} {\enquote {\bibinfo {title} {Charge nematicity
  and electronic raman scattering in iron-based superconductors},}\ }\href
  {\doibase 10.1016/j.crhy.2015.10.001} {\bibfield  {journal} {\bibinfo
  {journal} {C. R. Phys.}\ }\textbf {\bibinfo {volume} {17}},\ \bibinfo {pages}
  {113--139} (\bibinfo {year} {2016})}\BibitemShut {NoStop}%
\bibitem [{\citenamefont {Chu}\ \emph {et~al.}(2012)\citenamefont {Chu},
  \citenamefont {Kuo}, \citenamefont {Analytis},\ and\ \citenamefont
  {Fisher}}]{FisherScienceNematic2012}%
  \BibitemOpen
  \bibfield  {author} {\bibinfo {author} {\bibfnamefont {Jiun-Haw}\
  \bibnamefont {Chu}}, \bibinfo {author} {\bibfnamefont {Hsueh-Hui}\
  \bibnamefont {Kuo}}, \bibinfo {author} {\bibfnamefont {James~G.}\
  \bibnamefont {Analytis}}, \ and\ \bibinfo {author} {\bibfnamefont {Ian~R.}\
  \bibnamefont {Fisher}},\ }\bibfield  {title} {\enquote {\bibinfo {title}
  {Divergent nematic susceptibility in an iron arsenide superconductor},}\
  }\href {\doibase 10.1126/science.1221713} {\bibfield  {journal} {\bibinfo
  {journal} {Science}\ }\textbf {\bibinfo {volume} {337}},\ \bibinfo {pages}
  {710--712} (\bibinfo {year} {2012})}\BibitemShut {NoStop}%
\bibitem [{\citenamefont {Ni}\ \emph {et~al.}(2008)\citenamefont {Ni},
  \citenamefont {Tillman}, \citenamefont {Yan}, \citenamefont {Kracher},
  \citenamefont {Hannahs}, \citenamefont {Bud'ko},\ and\ \citenamefont
  {Canfield}}]{CanfieldBa122phasediagram2008}%
  \BibitemOpen
  \bibfield  {author} {\bibinfo {author} {\bibfnamefont {N.}~\bibnamefont
  {Ni}}, \bibinfo {author} {\bibfnamefont {M.~E.}\ \bibnamefont {Tillman}},
  \bibinfo {author} {\bibfnamefont {J.-Q.}\ \bibnamefont {Yan}}, \bibinfo
  {author} {\bibfnamefont {A.}~\bibnamefont {Kracher}}, \bibinfo {author}
  {\bibfnamefont {S.~T.}\ \bibnamefont {Hannahs}}, \bibinfo {author}
  {\bibfnamefont {S.~L.}\ \bibnamefont {Bud'ko}}, \ and\ \bibinfo {author}
  {\bibfnamefont {P.~C.}\ \bibnamefont {Canfield}},\ }\bibfield  {title}
  {\enquote {\bibinfo {title} {Effects of {Co} substitution on thermodynamic
  and transport properties and anisotropic {H$_{c2}$ in
  {Ba(Fe$_{1-x}$Co$_x$)$_2$As$_2$}} single crystals},}\ }\href {\doibase
  10.1103/PhysRevB.78.214515} {\bibfield  {journal} {\bibinfo  {journal} {Phys.
  Rev. B}\ }\textbf {\bibinfo {volume} {78}},\ \bibinfo {eid} {214515}
  (\bibinfo {year} {2008})}\BibitemShut {NoStop}%
\bibitem [{\citenamefont {Dioguardi}\ \emph {et~al.}(2016)\citenamefont
  {Dioguardi}, \citenamefont {Kissikov}, \citenamefont {Lin}, \citenamefont
  {Shirer}, \citenamefont {Lawson}, \citenamefont {Grafe}, \citenamefont {Chu},
  \citenamefont {Fisher}, \citenamefont {Fernandes},\ and\ \citenamefont
  {Curro}}]{DioguardiPdoped2015}%
  \BibitemOpen
  \bibfield  {author} {\bibinfo {author} {\bibfnamefont {A.~P.}\ \bibnamefont
  {Dioguardi}}, \bibinfo {author} {\bibfnamefont {T.}~\bibnamefont {Kissikov}},
  \bibinfo {author} {\bibfnamefont {C.~H.}\ \bibnamefont {Lin}}, \bibinfo
  {author} {\bibfnamefont {K.~R.}\ \bibnamefont {Shirer}}, \bibinfo {author}
  {\bibfnamefont {M.~M.}\ \bibnamefont {Lawson}}, \bibinfo {author}
  {\bibfnamefont {H.-J.}\ \bibnamefont {Grafe}}, \bibinfo {author}
  {\bibfnamefont {J.-H.}\ \bibnamefont {Chu}}, \bibinfo {author} {\bibfnamefont
  {I.~R.}\ \bibnamefont {Fisher}}, \bibinfo {author} {\bibfnamefont {R.~M.}\
  \bibnamefont {Fernandes}}, \ and\ \bibinfo {author} {\bibfnamefont {N.~J.}\
  \bibnamefont {Curro}},\ }\bibfield  {title} {\enquote {\bibinfo {title}
  {{NMR} evidence for inhomogeneous nematic fluctuations in
  {BaFe$_{2}$(As$_{1-x}$P$_{x}$)$_{2}$}},}\ }\href {\doibase
  10.1103/PhysRevLett.116.107202} {\bibfield  {journal} {\bibinfo  {journal}
  {Phys. Rev. Lett.}\ }\textbf {\bibinfo {volume} {116}},\ \bibinfo {pages}
  {107202} (\bibinfo {year} {2016})}\BibitemShut {NoStop}%
\bibitem [{\citenamefont {Iye}\ \emph {et~al.}(2015)\citenamefont {Iye},
  \citenamefont {Julien}, \citenamefont {Mayaffre}, \citenamefont
  {Horvati\"{c}}, \citenamefont {Berthier}, \citenamefont {Ishida},
  \citenamefont {Ikeda}, \citenamefont {Kasahara}, \citenamefont {Shibauchi},\
  and\ \citenamefont {Matsuda}}]{IyeJPSLorbitalnematicity2015}%
  \BibitemOpen
  \bibfield  {author} {\bibinfo {author} {\bibfnamefont {Tetsuya}\ \bibnamefont
  {Iye}}, \bibinfo {author} {\bibfnamefont {Marc-Henri}\ \bibnamefont
  {Julien}}, \bibinfo {author} {\bibfnamefont {Hadrien}\ \bibnamefont
  {Mayaffre}}, \bibinfo {author} {\bibfnamefont {Mladen}\ \bibnamefont
  {Horvati\"{c}}}, \bibinfo {author} {\bibfnamefont {Claude}\ \bibnamefont
  {Berthier}}, \bibinfo {author} {\bibfnamefont {Kenji}\ \bibnamefont
  {Ishida}}, \bibinfo {author} {\bibfnamefont {Hiroaki}\ \bibnamefont {Ikeda}},
  \bibinfo {author} {\bibfnamefont {Shigeru}\ \bibnamefont {Kasahara}},
  \bibinfo {author} {\bibfnamefont {Takasada}\ \bibnamefont {Shibauchi}}, \
  and\ \bibinfo {author} {\bibfnamefont {Yuji}\ \bibnamefont {Matsuda}},\
  }\bibfield  {title} {\enquote {\bibinfo {title} {Emergence of orbital
  nematicity in the tetragonal phase of {BaFe$_2$(As$_{1-x}$P$_x$)$_2$}},}\
  }\href {\doibase 10.7566/JPSJ.84.043705} {\bibfield  {journal} {\bibinfo
  {journal} {J. Phys. Soc. Jpn.}\ }\textbf {\bibinfo {volume} {84}},\ \bibinfo
  {pages} {043705} (\bibinfo {year} {2015})}\BibitemShut {NoStop}%
\bibitem [{\citenamefont {Ning}\ \emph {et~al.}(2009)\citenamefont {Ning},
  \citenamefont {Ahilan}, \citenamefont {Imai}, \citenamefont {Sefat},
  \citenamefont {Jin}, \citenamefont {McGuire}, \citenamefont {Sales},\ and\
  \citenamefont {Mandrus}}]{ImaiLightlyDoped}%
  \BibitemOpen
  \bibfield  {author} {\bibinfo {author} {\bibfnamefont {F.~L.}\ \bibnamefont
  {Ning}}, \bibinfo {author} {\bibfnamefont {K.}~\bibnamefont {Ahilan}},
  \bibinfo {author} {\bibfnamefont {T.}~\bibnamefont {Imai}}, \bibinfo {author}
  {\bibfnamefont {A.~S.}\ \bibnamefont {Sefat}}, \bibinfo {author}
  {\bibfnamefont {R.}~\bibnamefont {Jin}}, \bibinfo {author} {\bibfnamefont
  {M.~A.}\ \bibnamefont {McGuire}}, \bibinfo {author} {\bibfnamefont {B.~C.}\
  \bibnamefont {Sales}}, \ and\ \bibinfo {author} {\bibfnamefont
  {D.}~\bibnamefont {Mandrus}},\ }\bibfield  {title} {\enquote {\bibinfo
  {title} {{$^{59}$Co and $^{75}$As NMR investigation of lightly doped
  Ba(Fe$_{1-x}$Co$_x$)$_2$As$_2$ ($x$ = 0.02, 0.04)}},}\ }\href {\doibase
  10.1103/PhysRevB.79.140506} {\bibfield  {journal} {\bibinfo  {journal} {Phys.
  Rev. B}\ }\textbf {\bibinfo {volume} {79}},\ \bibinfo {pages} {140506}
  (\bibinfo {year} {2009})}\BibitemShut {NoStop}%
\bibitem [{\citenamefont {Dioguardi}\ \emph {et~al.}(2010)\citenamefont
  {Dioguardi}, \citenamefont {apRoberts Warren}, \citenamefont {Shockley},
  \citenamefont {Bud'ko}, \citenamefont {Ni}, \citenamefont {Canfield},\ and\
  \citenamefont {Curro}}]{Dioguardi2010}%
  \BibitemOpen
  \bibfield  {author} {\bibinfo {author} {\bibfnamefont {A.~P.}\ \bibnamefont
  {Dioguardi}}, \bibinfo {author} {\bibfnamefont {N.}~\bibnamefont {apRoberts
  Warren}}, \bibinfo {author} {\bibfnamefont {A.~C.}\ \bibnamefont {Shockley}},
  \bibinfo {author} {\bibfnamefont {S.~L.}\ \bibnamefont {Bud'ko}}, \bibinfo
  {author} {\bibfnamefont {N.}~\bibnamefont {Ni}}, \bibinfo {author}
  {\bibfnamefont {P.~C.}\ \bibnamefont {Canfield}}, \ and\ \bibinfo {author}
  {\bibfnamefont {N.~J.}\ \bibnamefont {Curro}},\ }\bibfield  {title} {\enquote
  {\bibinfo {title} {Local magnetic inhomogeneities in
  {Ba(Fe$_{1-x}$Ni$_x$)$_2$As$_2$} as seen via {As-75} {NMR}},}\ }\href
  {\doibase 10.1103/PhysRevB.82.140411} {\bibfield  {journal} {\bibinfo
  {journal} {Phys. Rev. B}\ }\textbf {\bibinfo {volume} {82}},\ \bibinfo
  {pages} {140411(R)} (\bibinfo {year} {2010})}\BibitemShut {NoStop}%
\bibitem [{\citenamefont {Takeda}\ \emph {et~al.}(2014)\citenamefont {Takeda},
  \citenamefont {Imai}, \citenamefont {Tachibana}, \citenamefont {Gaudet},
  \citenamefont {Gaulin}, \citenamefont {Saparov},\ and\ \citenamefont
  {Sefat}}]{Takeda:2014ia}%
  \BibitemOpen
  \bibfield  {author} {\bibinfo {author} {\bibfnamefont {Hikaru}\ \bibnamefont
  {Takeda}}, \bibinfo {author} {\bibfnamefont {Takashi}\ \bibnamefont {Imai}},
  \bibinfo {author} {\bibfnamefont {Makoto}\ \bibnamefont {Tachibana}},
  \bibinfo {author} {\bibfnamefont {Jonathan}\ \bibnamefont {Gaudet}}, \bibinfo
  {author} {\bibfnamefont {Bruce~D.}\ \bibnamefont {Gaulin}}, \bibinfo {author}
  {\bibfnamefont {Bayrammurad~I.}\ \bibnamefont {Saparov}}, \ and\ \bibinfo
  {author} {\bibfnamefont {Athena~S.}\ \bibnamefont {Sefat}},\ }\bibfield
  {title} {\enquote {\bibinfo {title} {{Cu} substitution effects on the local
  magnetic properties of {Ba(Fe$_{1-x}$Cu$_x$)$_2$As$_2$}: A site-selective
  $^{75}${As} and $^{63}${Cu} {NMR} study},}\ }\href {\doibase
  10.1103/PhysRevLett.113.117001} {\bibfield  {journal} {\bibinfo  {journal}
  {Phys. Rev. Lett.}\ }\textbf {\bibinfo {volume} {113}},\ \bibinfo {pages}
  {117001} (\bibinfo {year} {2014})}\BibitemShut {NoStop}%
\bibitem [{\citenamefont {Baek}\ \emph {et~al.}(2015)\citenamefont {Baek},
  \citenamefont {Efremov}, \citenamefont {Ok}, \citenamefont {Kim},
  \citenamefont {van~den Brink},\ and\ \citenamefont {B\"{u}chner}}]{Baek2015}%
  \BibitemOpen
  \bibfield  {author} {\bibinfo {author} {\bibfnamefont {S-H.}\ \bibnamefont
  {Baek}}, \bibinfo {author} {\bibfnamefont {D.~V.}\ \bibnamefont {Efremov}},
  \bibinfo {author} {\bibfnamefont {J.~M.}\ \bibnamefont {Ok}}, \bibinfo
  {author} {\bibfnamefont {J.~S.}\ \bibnamefont {Kim}}, \bibinfo {author}
  {\bibfnamefont {Jeroen}\ \bibnamefont {van~den Brink}}, \ and\ \bibinfo
  {author} {\bibfnamefont {B.}~\bibnamefont {B\"{u}chner}},\ }\bibfield
  {title} {\enquote {\bibinfo {title} {Orbital-driven nematicity in {FeSe}},}\
  }\href {http://dx.doi.org/10.1038/nmat4138} {\bibfield  {journal} {\bibinfo
  {journal} {Nat. Mater.}\ }\textbf {\bibinfo {volume} {14}},\ \bibinfo {pages}
  {210--214} (\bibinfo {year} {2015})}\BibitemShut {NoStop}%
\bibitem [{\citenamefont {He}\ \emph {et~al.}()\citenamefont {He},
  \citenamefont {Wang}, \citenamefont {Ahn}, \citenamefont {Hardy},
  \citenamefont {Wolf}, \citenamefont {Adelmann}, \citenamefont {Schmalian},
  \citenamefont {Eremin},\ and\ \citenamefont
  {Meingast}}]{MeingastBaFe2As2strain2016}%
  \BibitemOpen
  \bibfield  {author} {\bibinfo {author} {\bibfnamefont {Mingquan}\
  \bibnamefont {He}}, \bibinfo {author} {\bibfnamefont {Liran}\ \bibnamefont
  {Wang}}, \bibinfo {author} {\bibfnamefont {Felix}\ \bibnamefont {Ahn}},
  \bibinfo {author} {\bibfnamefont {Frédéric}\ \bibnamefont {Hardy}}, \bibinfo
  {author} {\bibfnamefont {Thomas}\ \bibnamefont {Wolf}}, \bibinfo {author}
  {\bibfnamefont {Peter}\ \bibnamefont {Adelmann}}, \bibinfo {author}
  {\bibfnamefont {Jörg}\ \bibnamefont {Schmalian}}, \bibinfo {author}
  {\bibfnamefont {Ilya}\ \bibnamefont {Eremin}}, \ and\ \bibinfo {author}
  {\bibfnamefont {Christoph}\ \bibnamefont {Meingast}},\ }\bibfield  {title}
  {\enquote {\bibinfo {title} {Dichotomy between in-plane magnetic
  susceptibility and resistivity anisotropies in extremely strained
  {BaFe$_{2}$As$_{2}$}},}\ }\href@noop {} {\ }\Eprint
  {http://arxiv.org/abs/1610.05575v2} {1610.05575v2} \BibitemShut {NoStop}%
\bibitem [{\citenamefont {Christensen}\ \emph {et~al.}(2015)\citenamefont
  {Christensen}, \citenamefont {Kang}, \citenamefont {Andersen}, \citenamefont
  {Eremin},\ and\ \citenamefont {Fernandes}}]{Christensen2015}%
  \BibitemOpen
  \bibfield  {author} {\bibinfo {author} {\bibfnamefont {Morten~H.}\
  \bibnamefont {Christensen}}, \bibinfo {author} {\bibfnamefont {Jian}\
  \bibnamefont {Kang}}, \bibinfo {author} {\bibfnamefont {Brian~M.}\
  \bibnamefont {Andersen}}, \bibinfo {author} {\bibfnamefont {Ilya}\
  \bibnamefont {Eremin}}, \ and\ \bibinfo {author} {\bibfnamefont {Rafael~M.}\
  \bibnamefont {Fernandes}},\ }\bibfield  {title} {\enquote {\bibinfo {title}
  {Spin reorientation driven by the interplay between spin-orbit coupling and
  hund's rule coupling in iron pnictides},}\ }\href {\doibase
  10.1103/physrevb.92.214509} {\bibfield  {journal} {\bibinfo  {journal} {Phys.
  Rev. B}\ }\textbf {\bibinfo {volume} {92}},\ \bibinfo {pages} {214509}
  (\bibinfo {year} {2015})}\BibitemShut {NoStop}%
\end{thebibliography}%

\section{Spectral Measurements}

When the crystal is strained by applying voltage to the piezoelectric
stacks, the displacement, $x$, is measured by a capacitive position
sensor, and strain is calculated as $\epsilon=(x-x_{0})/L_{0}$, where
$L_{0}$ is the unstrained length of the crystal. It is crucial to
determine the unstrained displacement, $x_{0}$, at cryogenic temperatures
due to differential thermal contraction between the strain device
and the sample. This value can be obtained by observing the asymmetry
of the electric field gradient (EFG) tensor. The spectra were measured
by acquiring echoes while sweeping the magnetic field $H_{0}$ at
fixed frequency. The quadrupolar satellite resonances occur at fields
$H_{sat}=(f_{0}\pm\nu_{\alpha\alpha})/\gamma(1+K_{\alpha\alpha})$,
where $f_{0}$ is the radiofrequency, $\gamma=7.29019$ MHz/T is the
gyromagnetic ratio, $K_{\alpha\alpha}$ and $\nu_{\alpha\alpha}$
are the Knight shift and EFG tensor components in the $\alpha=(x,y,z)$
direction. The central transition field is given by: $H_{cen}=\frac{f_{0}}{\gamma(1+K_{\alpha\alpha})}\left(\frac{1}{2}+\sqrt{\frac{3f_{0}^{2}-2(\nu_{\beta\beta}+\nu_{\alpha\alpha})^{2}}{12}}\right)$,
where $\beta=(y,x,z)$ for $\alpha={x,y,z}$. The center of gravity
of each peak was used to determine the resonance field, and hence
$K_{\alpha\alpha}$ and $\nu_{\alpha\alpha}$ as a function of strain.

\begin{figure}[!tb]
\includegraphics[width=\linewidth]{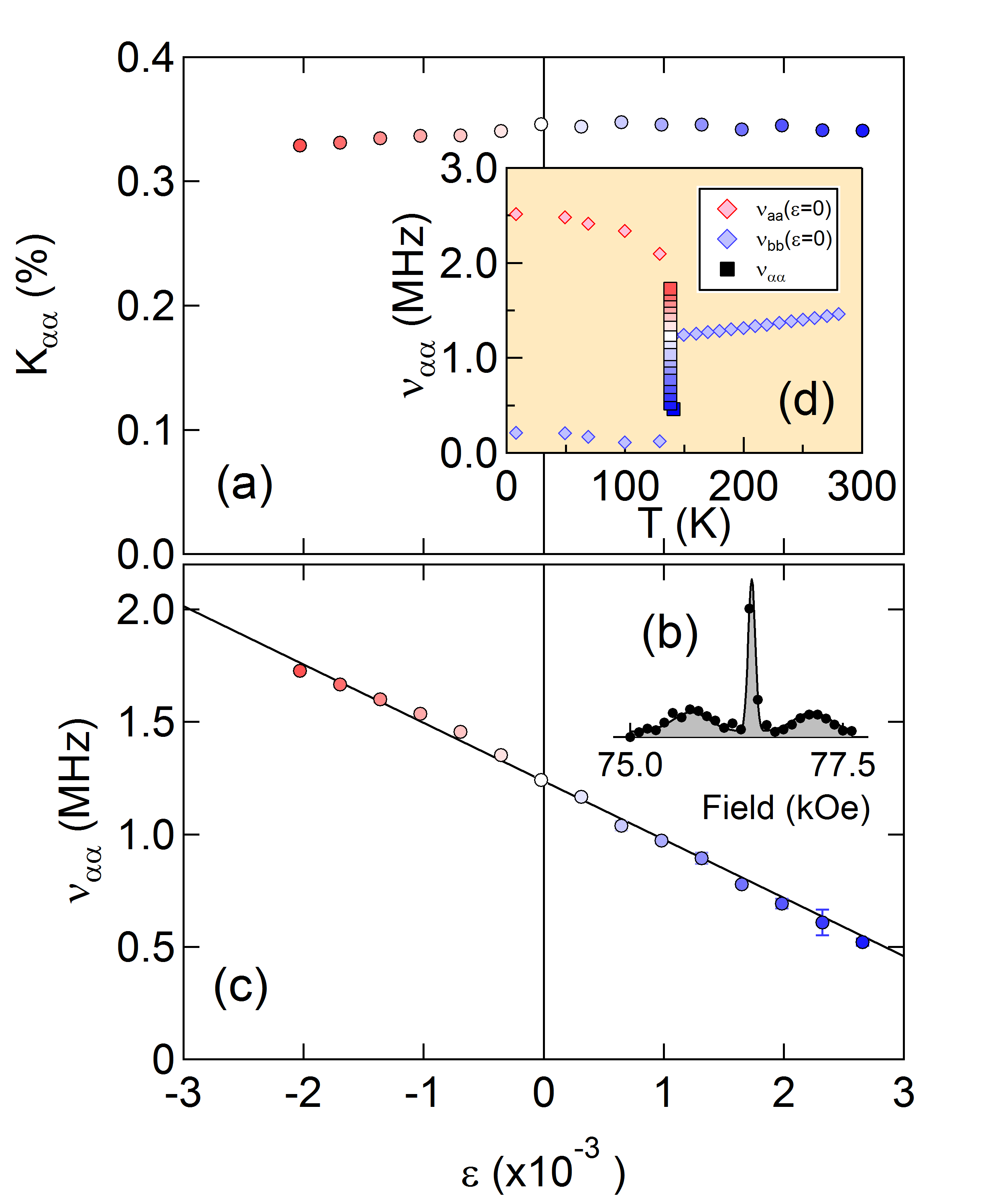}
\centering \caption{\label{fig:EFGandKS} (a) Knight shift versus strain at 138K. (b)
The $^{75}$As spectrum at 138K for a strain level of 0.0265\% at
frequency 55.924 MHz. The solid line is a fit to the spectrum as described
in the text. (c) The quadrupolar splitting versus strain, and (d)
versus temperature. The zero-strain points (diamonds) are reproduced
from Ref. \protect\onlinecite{takigawa2008}.}
\end{figure}

Fig. \ref{fig:EFGandKS}(b) shows a typical field-swept NMR spectrum
of the $^{75}$As, revealing a narrow central transition ($I_{z}=1/2\leftrightarrow-1/2$)
and two quadrupolar satellite peaks ($\pm3/2\leftrightarrow\pm1/2$).
The spectrum was fit to the sum of three Gaussians to extract both
the Knight shift, $K_{\alpha\alpha}$, and the EFG, $\nu_{\alpha\alpha}$.
The EFG tensor is given by $\nu_{\alpha\beta}=(eQ/12h)\partial^{2}V/\partial x_{\alpha}\partial x_{\beta}$,
where $Q$ is the quadrupolar moment of the $^{75}$As and $V$ is
the electrostatic potential at the As site. This quantity is dominated
by the occupation of the $d_{xz,yz}$-orbitals of the neighboring
Fe atoms, and the EFG asymmetry $\eta=(\nu_{yy}-\nu_{xx})/(\nu_{xx}+\nu_{yy})$
is a measure of the nematic order parameter \cite{DioguardiPdoped2015,IyeJPSLorbitalnematicity2015}.
Note that the magnetic field lies along the shorter $b$-axis under
tensile strain ($\epsilon>0$), and along the longer $a$-axis under
compressive strain ($\epsilon<0$), as shown in Fig. 2 of the main
text. The EFG enables us to identify the zero-strain displacement,
$x_{0}$, by the condition $|\nu_{xx}|=|\nu_{yy}|=|\nu_{zz}|/2$.
As shown in Fig. \ref{fig:EFGandKS}(c), $\nu_{yy}$, and hence $\eta(\epsilon)=(\nu_{yy}(\epsilon)-\nu_{yy}(-\epsilon))/(\nu_{yy}(\epsilon)+\nu_{yy}(-\epsilon))$,
varies linearly with strain.

Despite the fact that the EFG varies with strain, we find no significant
variation of the satellite linewidth with strain. The strong variation
of the EFG with strain explains the quadrupolar broadening observed
in Co, Ni or Cu-doped Ba(Fe,M)$_{2}$As$_{2}$ \cite{ImaiLightlyDoped,Dioguardi2010,Takeda:2014ia}.
The dopant atoms create an inhomogeneous strain field that gives rise
to a distribution of local EFGs. Recently a finite value of $\eta\sim0.1$
was reported in the tetragonal phase of unstrained BaFe$_{2}$(As$_{1-x}$P$_{x}$)$_{2}$
above $T_{s}$ \cite{IyeJPSLorbitalnematicity2015}. The origin of
this finite nematicity is likely due to local defects, and based on
our results the strain fields are on the order of $0.05\%$.

The Knight shift is shown versus strain in Fig. \ref{fig:EFGandKS}(a)
for $\mathbf{H}_{0}\perp c$. The in-plane Knight shift shows little
or no variation with $\epsilon$, such that $(K_{xx}-K_{yy})/K_{yy}\leq3\%$
at the highest strain levels in this material. This result is surprising
because the same quantity is approximately 6\% in the nematic phase
of FeSe \cite{Baek2015}. Recent static susceptibility measurements
in BaFe$_{2}$As$_{2}$ under strain indicate that $\chi_{xx}$ and
$\chi_{yy}$ in the paramagnetic phase differ by only 5\% \cite{MeingastBaFe2As2strain2016}.
This result suggests that $\chi_{\alpha\alpha}(\mathbf{q}=0)$ couples
only weakly to the strain.

\begin{widetext}

\section{Spin-Lattice Relaxation Rate: Model}

As stated in the main text, the spin-lattice relaxation rate is given
by:

\begin{equation}
\left(\frac{1}{T_{1}T}\right)_{\mu}=\frac{\gamma^{2}}{2}\sum\limits _{\mathbf{q},\alpha,\beta}\mathcal{F}_{\alpha\beta}^{(\mu)}(\mathbf{q})\frac{\textrm{Im}\chi_{\alpha\beta}(\mathbf{q},\omega)}{\hslash\omega}
\end{equation}
where $\gamma$ is the gyromagnetic ratio of the nuclear spin, and
$\mathcal{F}_{\alpha\beta}^{(\mu)}$ is a form factor that depends
on the direction of the applied field (indicated by $\mu$), and $\alpha,\beta=\left\{ x,y,z\right\} $.
The coordinate system is defined such that $x$ and $y$ connect nearest
neighbor Fe atoms. Ref. \cite{T1formfactorsArsenides} derived the
form factor for an As nucleus subject to an arbitrary field direction.
In the paramagnetic state, one obtains (see also Ref. \cite{NMRnematicStrainBa122PRB2016}):

\begin{equation}
\left(\frac{1}{T_{1}T}\right)_{\mu}=\frac{\gamma^{2}}{2}\sum_{\mathbf{q}}\sum_{\alpha=1,2}\left[\bar{R}^{(\mu)}\cdot\bar{A}_{\mathbf{q}}\cdot\bar{\tilde{\chi}}\left(\mathbf{q}\right)\cdot\bar{A}_{\mathbf{q}}^{\dagger}\cdot\left(\bar{R}^{(\mu)}\right)^{\dagger}\right]_{\alpha\alpha}\label{def}
\end{equation}

All quantities with an overbar are $3\times3$ matrices. The matrix
$\bar{\tilde{\chi}}\left(\mathbf{q}\right)$ is diagonal; its matrix
elements are related to the magnetic susceptibility elements according
to:

\begin{equation}
\tilde{\chi}_{\alpha\alpha}\left(\mathbf{q}\right)\equiv\lim_{\omega\rightarrow0}\frac{\mathrm{Im}\chi_{\alpha\alpha}\left(\mathbf{q},\omega\right)}{\hbar\omega}=\frac{1}{\Gamma}\chi_{\alpha\alpha}^{2}(\mathbf{q})\label{aux_def}
\end{equation}
where $\Gamma$ is the Landau damping term. Furthermore, we have the
hyperfine tensor:

\begin{equation}
\bar{A}_{\mathbf{q}}=4\left(\begin{array}{ccc}
A_{xx}\cos\left(\frac{q_{x}}{2}\right)\cos\left(\frac{q_{y}}{2}\right) & -A_{xy}\sin\left(\frac{q_{x}}{2}\right)\sin\left(\frac{q_{y}}{2}\right) & iA_{xz}\sin\left(\frac{q_{x}}{2}\right)\cos\left(\frac{q_{y}}{2}\right)\\
-A_{yx}\sin\left(\frac{q_{x}}{2}\right)\sin\left(\frac{q_{y}}{2}\right) & A_{yy}\cos\left(\frac{q_{x}}{2}\right)\cos\left(\frac{q_{y}}{2}\right) & iA_{yz}\cos\left(\frac{q_{x}}{2}\right)\sin\left(\frac{q_{y}}{2}\right)\\
iA_{zx}\sin\left(\frac{q_{x}}{2}\right)\cos\left(\frac{q_{y}}{2}\right) & iA_{zy}\cos\left(\frac{q_{x}}{2}\right)\sin\left(\frac{q_{y}}{2}\right) & A_{zz}\cos\left(\frac{q_{x}}{2}\right)\cos\left(\frac{q_{y}}{2}\right)
\end{array}\right)
\end{equation}
and the rotation matrix:

\begin{equation}
\bar{R}^{(\mu)}=\left(\begin{array}{ccc}
\sin^{2}\phi+\cos\theta\,\cos^{2}\phi & -\sin2\phi\,\sin^{2}\frac{\theta}{2} & \cos\phi\,\sin\theta\\
-\sin2\phi\,\sin^{2}\frac{\theta}{2} & \cos^{2}\phi+\cos\theta\,\sin^{2}\phi & \sin\phi\,\sin\theta\\
-\cos\phi\,\sin\theta & -\sin\phi\,\sin\theta & \cos\theta
\end{array}\right)
\end{equation}

Here, the field direction $\mu$ is described by the angles $\theta,\varphi$
according to $\hat{\mathbf{h}}=\cos\varphi\sin\theta\,\hat{\mathbf{x}}+\sin\varphi\sin\theta\,\hat{\mathbf{y}}+\cos\theta\,\hat{\mathbf{z}}$.
Because the lattice distortion is very small, we consider hereafter
that the hyperfine tensor remains essentially tetragonal, i.e. $A_{xx}=A_{yy}$,
$A_{yz}=A_{xz}$, and $A_{xy}=A_{yx}$.

It is now straightforward to obtain the expressions for $1/\left(T_{1}T\right)_{\mu}$
for different field directions $\mu$. We find:

\begin{eqnarray}
\left(\frac{1}{T_{1}T}\right)_{x} & = & 8\gamma^{2}\sum_{\mathbf{q}}\left[\sin^{2}\left(\frac{q_{x}}{2}\right)\sin^{2}\left(\frac{q_{y}}{2}\right)A_{xy}^{2}+\sin^{2}\left(\frac{q_{x}}{2}\right)\cos^{2}\left(\frac{q_{y}}{2}\right)A_{xz}^{2}\right]\tilde{\chi}_{xx}\left(\mathbf{q}\right)\nonumber \\
 &  & 8\gamma^{2}\sum_{\mathbf{q}}\left[\cos^{2}\left(\frac{q_{x}}{2}\right)\cos^{2}\left(\frac{q_{y}}{2}\right)A_{yy}^{2}+\cos^{2}\left(\frac{q_{x}}{2}\right)\sin^{2}\left(\frac{q_{y}}{2}\right)A_{yz}^{2}\right]\tilde{\chi}_{yy}\left(\mathbf{q}\right)\nonumber \\
 &  & 8\gamma^{2}\sum_{\mathbf{q}}\left[\cos^{2}\left(\frac{q_{x}}{2}\right)\sin^{2}\left(\frac{q_{y}}{2}\right)A_{yz}^{2}+\cos^{2}\left(\frac{q_{x}}{2}\right)\cos^{2}\left(\frac{q_{y}}{2}\right)A_{zz}^{2}\right]\tilde{\chi}_{zz}\left(\mathbf{q}\right)\label{T1T_a_recap}
\end{eqnarray}

\begin{eqnarray}
\left(\frac{1}{T_{1}T}\right)_{y} & = & 8\gamma^{2}\sum_{\mathbf{q}}\left[\cos^{2}\left(\frac{q_{x}}{2}\right)\cos^{2}\left(\frac{q_{y}}{2}\right)A_{xx}^{2}+\sin^{2}\left(\frac{q_{x}}{2}\right)\cos^{2}\left(\frac{q_{y}}{2}\right)A_{xz}^{2}\right]\tilde{\chi}_{xx}\left(\mathbf{q}\right)\nonumber \\
 &  & 8\gamma^{2}\sum_{\mathbf{q}}\left[\sin^{2}\left(\frac{q_{x}}{2}\right)\sin^{2}\left(\frac{q_{y}}{2}\right)A_{xy}^{2}+\cos^{2}\left(\frac{q_{x}}{2}\right)\sin^{2}\left(\frac{q_{y}}{2}\right)A_{yz}^{2}\right]\tilde{\chi}_{yy}\left(\mathbf{q}\right)\nonumber \\
 &  & 8\gamma^{2}\sum_{\mathbf{q}}\left[\sin^{2}\left(\frac{q_{x}}{2}\right)\cos^{2}\left(\frac{q_{y}}{2}\right)A_{xz}^{2}+\cos^{2}\left(\frac{q_{x}}{2}\right)\cos^{2}\left(\frac{q_{y}}{2}\right)A_{zz}^{2}\right]\tilde{\chi}_{zz}\left(\mathbf{q}\right)\label{T1T_b_recap}
\end{eqnarray}
and:

\begin{align}
\left(\frac{1}{T_{1}T}\right)_{z} & =8\gamma^{2}\sum_{\mathbf{q}}\left[\cos^{2}\left(\frac{q_{x}}{2}\right)\cos^{2}\left(\frac{q_{y}}{2}\right)A_{xx}^{2}+\sin^{2}\left(\frac{q_{x}}{2}\right)\sin^{2}\left(\frac{q_{y}}{2}\right)A_{xy}^{2}\right]\tilde{\chi}_{xx}\left(\mathbf{q}\right)\nonumber \\
 & 8\gamma^{2}\sum_{\mathbf{q}}\left[\cos^{2}\left(\frac{q_{x}}{2}\right)\cos^{2}\left(\frac{q_{y}}{2}\right)A_{yy}^{2}+\sin^{2}\left(\frac{q_{x}}{2}\right)\sin^{2}\left(\frac{q_{y}}{2}\right)A_{xy}^{2}\right]\tilde{\chi}_{yy}\left(\mathbf{q}\right)\nonumber \\
 & 8\gamma^{2}\sum_{\mathbf{q}}\left[\sin^{2}\left(\frac{q_{x}}{2}\right)\cos^{2}\left(\frac{q_{y}}{2}\right)A_{xz}^{2}+\cos^{2}\left(\frac{q_{x}}{2}\right)\sin^{2}\left(\frac{q_{y}}{2}\right)A_{yz}^{2}\right]\tilde{\chi}_{zz}\left(\mathbf{q}\right)\label{T1T_c_recap}
\end{align}

If we approximate the magnetic susceptibility as delta-functions peaked
at the magnetic ordering vectors $\mathbf{Q}_{1}=\left(\pi,0\right)$
and $\mathbf{Q}_{2}=\left(0,\pi\right)$, we obtain:

\begin{eqnarray}
(T_{1}T)_{x}^{-1} & = & \frac{8\gamma^{2}A_{xz}^{2}}{\Gamma}\left[\chi_{xx}^{2}\left(\mathbf{Q}_{1}\right)+\chi_{yy}^{2}\left(\mathbf{Q}_{2}\right)+\chi_{zz}^{2}\left(\mathbf{Q}_{2}\right)\right]\\
(T_{1}T)_{y}^{-1} & = & \frac{8\gamma^{2}A_{xz}^{2}}{\Gamma}\left[\chi_{xx}^{2}\left(\mathbf{Q}_{1}\right)+\chi_{yy}^{2}\left(\mathbf{Q}_{2}\right)+\chi_{zz}^{2}\left(\mathbf{Q}_{1}\right)\right]\\
(T_{1}T)_{z}^{-1} & = & \frac{8\gamma^{2}A_{xz}^{2}}{\Gamma}\left[\chi_{zz}^{2}\left(\mathbf{Q}_{1}\right)+\chi_{zz}^{2}\left(\mathbf{Q}_{2}\right)\right]
\end{eqnarray}

These equations can be inverted to extract the quantities:
\begin{eqnarray}
\chi_{zz}^{2}\left(\mathbf{Q}_{1}\right) & = & \frac{\Gamma}{16\gamma^{2}A_{xz}^{2}}\left[-(T_{1}T)_{y}^{-1}(-\epsilon)+(T_{1}T)_{y}^{-1}(\epsilon)+(T_{1}T)_{z}^{-1}(\epsilon)\right]\label{invert1}\\
\chi_{zz}^{2}\left(\mathbf{Q}_{2}\right) & = & \frac{\Gamma}{16\gamma^{2}A_{xz}^{2}}\left[(T_{1}T)_{y}^{-1}(-\epsilon)-(T_{1}T)_{y}^{-1}(\epsilon)+(T_{1}T)_{z}^{-1}(\epsilon)\right]\label{invert2}\\
\chi_{xx}^{2}\left(\mathbf{Q}_{1}\right)+\chi_{yy}^{2}\left(\mathbf{Q}_{2}\right) & = & \frac{\Gamma}{16\gamma^{2}A_{xz}^{2}}\left[(T_{1}T)_{y}^{-1}(-\epsilon)+(T_{1}T)_{y}^{-1}(\epsilon)-(T_{1}T)_{z}^{-1}(\epsilon)\right],\label{invert3}
\end{eqnarray}

using the fact that $(T_{1}T)_{x}^{-1}(\epsilon)=(T_{1}T)_{y}^{-1}(-\epsilon)$.
These quantities are plotted in Fig. 3(e) of the main text.

Although useful for a qualitative analysis, this approximation neglects
the important fact that the magnetic fluctuations have finite correlation
lengths $\xi$. To model this effect, we consider susceptibilities
peaked at $\mathbf{Q}_{1}$ and $\mathbf{Q}_{2}$, as seen by neutron
scattering experiments (the amplitude $\chi_{0}$ of the susceptibilities
is absorbed in $\Gamma$, for convenience) \cite{StrainedPnictidesNS2014science}:

\begin{eqnarray}
\Gamma\tilde{\chi}_{xx}\left(\mathbf{q}\right) & = & \frac{1}{\left[\left(\xi_{x}^{-2}-\varphi_{xy}\right)+\left(\cos q_{x}-\cos q_{y}+2\right)\right]^{2}}+\frac{1}{\left[\left(\xi_{y}^{-2}+\varphi_{yx}\right)+\left(-\cos q_{x}+\cos q_{y}+2\right)\right]^{2}}\nonumber \\
\Gamma\tilde{\chi}_{yy}\left(\mathbf{q}\right) & = & \frac{1}{\left[\left(\xi_{y}^{-2}-\varphi_{yx}\right)+\left(\cos q_{x}-\cos q_{y}+2\right)\right]^{2}}+\frac{1}{\left[\left(\xi_{x}^{-2}+\varphi_{xy}\right)+\left(-\cos q_{x}+\cos q_{y}+2\right)\right]^{2}}\nonumber \\
\Gamma\tilde{\chi}_{zz}\left(\mathbf{q}\right) & = & \frac{1}{\left[\left(\xi_{y}^{-2}-\varphi_{zz}\right)+\left(\cos q_{x}-\cos q_{y}+2\right)\right]^{2}}+\frac{1}{\left[\left(\xi_{y}^{-2}+\varphi_{zz}\right)+\left(-\cos q_{x}+\cos q_{y}+2\right)\right]^{2}},\nonumber\\\label{chi_def}
\end{eqnarray}

Note that we have three different correlation lengths: $\xi_{x}$
corresponds to in-plane spin fluctuations with spins parallel to the
ordering vector direction; $\xi_{y}$ corresponds to in-plane spin
fluctuations with spins perpendicular to the ordering vector direction;
and $\xi_{z}$ corresponds to out-of-plane spin fluctuations. This
spin anisotropy originates from the spin-orbit coupling, as shown
in Ref. \cite{Christensen2015}. The nematic order parameters $\varphi_{\alpha\beta}$
split the tetragonal degeneracy between $\chi_{xx}\left(\mathbf{Q}_{1}\right)$
and $\chi_{yy}(\mathbf{Q}_{2})$, between $\chi_{xx}\left(\mathbf{Q}_{2}\right)$
and $\chi_{yy}(\mathbf{Q}_{1})$, and between $\chi_{zz}\left(\mathbf{Q}_{1}\right)$
and $\chi_{zz}(\mathbf{Q}_{2})$. They are related to the external
strain $\epsilon$ according to the nematic susceptibilities $\chi_{\mathrm{nem}}^{(\alpha\beta)}$,
i.e. $\varphi_{\alpha\beta}=\epsilon\chi_{\mathrm{nem}}^{(\alpha\beta)}$.

Substituting these expressions in Eqs. (\ref{T1T_a_recap}), (\ref{T1T_b_recap}),
and (\ref{T1T_c_recap}) give:

\begin{eqnarray}
\frac{\Gamma}{8\gamma^{2}}\left(\frac{1}{T_{1}T}\right)_{x} & = & A_{xy}^{2}\left[J_{1}\left(\xi_{x}^{-2}-\varphi_{xy}\right)+J_{1}\left(\xi_{y}^{-2}+\varphi_{yx}\right)\right]+A_{xz}^{2}\left[J_{3}\left(\xi_{x}^{-2}-\varphi_{xy}\right)+J_{2}\left(\xi_{y}^{-2}+\varphi_{yx}\right)\right]\nonumber \\
 &  & +A_{yy}^{2}\left[J_{1}\left(\xi_{y}^{-2}-\varphi_{yx}\right)+J_{1}\left(\xi_{x}^{-2}+\varphi_{xy}\right)\right]+A_{yz}^{2}\left[J_{2}\left(\xi_{y}^{-2}-\varphi_{yx}\right)+J_{3}\left(\xi_{x}^{-2}+\varphi_{xy}\right)\right]\nonumber \\
 &  & +A_{yz}^{2}\left[J_{2}\left(\xi_{z}^{-2}-\varphi_{zz}\right)+
J_{3}\left(\xi_{z}^{-2}+\varphi_{zz}\right)\right]+A_{zz}^{2}\left[J_{1}\left(\xi_{z}^{-2}-\varphi_{zz}\right)+J_{1}\left(\xi_{z}^{-2}+\varphi_{zz}\right)\right]\nonumber \\
\label{eq1}
\end{eqnarray}
as well as

\begin{eqnarray}
\frac{\Gamma}{8\gamma^{2}}\left(\frac{1}{T_{1}T}\right)_{y} & = & A_{xx}^{2}\left[J_{1}\left(\xi_{x}^{-2}-\varphi_{xy}\right)+J_{1}\left(\xi_{y}^{-2}+\varphi_{yx}\right)\right]+A_{xz}^{2}\left[J_{3}\left(\xi_{x}^{-2}-\varphi_{xy}\right)+J_{2}\left(\xi_{y}^{-2}+\varphi_{yx}\right)\right]\nonumber \\
 &  & +A_{xy}^{2}\left[J_{1}\left(\xi_{y}^{-2}-\varphi_{yx}\right)+J_{1}\left(\xi_{x}^{-2}+\varphi_{xy}\right)\right]+A_{yz}^{2}\left[J_{2}\left(\xi_{y}^{-2}-\varphi_{yx}\right)+J_{3}\left(\xi_{x}^{-2}+\varphi_{xy}\right)\right]\nonumber \\
 &  & +A_{xz}^{2}\left[J_{3}\left(\xi_{z}^{-2}-\varphi_{zz}\right)+J_{2}\left(\xi_{z}^{-2}+\varphi_{zz}\right)\right]+A_{zz}^{2}\left[J_{1}\left(\xi_{z}^{-2}-\varphi_{zz}\right)+J_{1}\left(\xi_{z}^{-2}+\varphi_{zz}\right)\right]\nonumber \\ \label{eq2}
\end{eqnarray}
and

\begin{align}
\frac{\Gamma}{8\gamma^{2}}\left(\frac{1}{T_{1}T}\right)_{z} & =A_{xx}^{2}\left[J_{1}\left(\xi_{x}^{-2}-\varphi_{xy}\right)+J_{1}\left(\xi_{y}^{-2}+\varphi_{yx}\right)\right]+A_{xy}^{2}\left[J_{1}\left(\xi_{x}^{-2}-\varphi_{xy}\right)+J_{1}\left(\xi_{y}^{-2}+\varphi_{yx}\right)\right]\nonumber \\
 & +A_{yy}^{2}\left[J_{1}\left(\xi_{y}^{-2}-\varphi_{yx}\right)+J_{1}\left(\xi_{x}^{-2}+\varphi_{xy}\right)\right]+A_{xy}^{2}\left[J_{1}\left(\xi_{y}^{-2}-\varphi_{yx}\right)+J_{1}\left(\xi_{x}^{-2}+\varphi_{xy}\right)\right]\nonumber \\
 & +A_{xz}^{2}\left[J_{3}\left(\xi_{z}^{-2}-\varphi_{zz}\right)+J_{2}\left(\xi_{z}^{-2}+\varphi_{zz}\right)\right]+A_{yz}^{2}\left[J_{2}\left(\xi_{z}^{-2}-\varphi_{zz}\right)+J_{3}\left(\xi_{z}^{-2}+\varphi_{zz}\right)\right]\nonumber \\ \label{eq3}
\end{align}

Here, we defined the integrals:

\begin{eqnarray}
J_{1}\left(r\right) & = & \int_{-\pi}^{\pi}\int_{-\pi}^{\pi}\frac{dq_{x}dq_{y}}{\left(2\pi\right)^{2}}\,\frac{\cos^{2}\left(\frac{q_{x}}{2}\right)\cos^{2}\left(\frac{q_{y}}{2}\right)}{\left[r+\left(\cos q_{x}-\cos q_{y}+2\right)\right]^{2}}\equiv\int_{-\pi}^{\pi}\int_{-\pi}^{\pi}\frac{dq_{x}dq_{y}}{\left(2\pi\right)^{2}}\,\frac{\sin^{2}\left(\frac{q_{x}}{2}\right)\sin^{2}\left(\frac{q_{y}}{2}\right)}{\left[r+\left(\cos q_{x}-\cos q_{y}+2\right)\right]^{2}}\nonumber \\
J_{2}\left(r\right) & = & \int_{-\pi}^{\pi}\int_{-\pi}^{\pi}\frac{dq_{x}dq_{y}}{\left(2\pi\right)^{2}}\,\frac{\cos^{2}\left(\frac{q_{x}}{2}\right)\sin^{2}\left(\frac{q_{y}}{2}\right)}{\left[r+\left(\cos q_{x}-\cos q_{y}+2\right)\right]^{2}}\nonumber \\
J_{3}\left(r\right) & = & \int_{-\pi}^{\pi}\int_{-\pi}^{\pi}\frac{dq_{x}dq_{y}}{\left(2\pi\right)^{2}}\,\frac{\sin^{2}\left(\frac{q_{x}}{2}\right)\cos^{2}\left(\frac{q_{y}}{2}\right)}{\left[r+\left(\cos q_{x}-\cos q_{y}+2\right)\right]^{2}}\label{integrals}
\end{eqnarray}

In the limit $\xi_{i}^{-2}\pm\varphi_{\alpha\beta}\ll1$, we can approximate
the integrals by expanding the integrand near $\left(\pi,0\right)$,
yielding:

\begin{align}
J_{1}\left(r\right) & \approx\frac{1}{4\pi}\ln\left(\frac{\Lambda_{1}}{\sqrt{r}}\right)\nonumber \\
J_{2}\left(r\right) & \approx\frac{1}{8\pi}\left[1-\frac{r}{2}\,\ln\left(\frac{\Lambda_{2}}{\sqrt{r}}\right)\right]\nonumber \\
J_{3}\left(r\right) & \approx\frac{1}{2\pi r}
\end{align}

where $\Lambda_{1}\approx1.45$ and $\Lambda_{2}\approx3.2$ for $r<0.5$,
according to numerical evaluations of the integrals. Note that, as
expected from symmetry considerations, $\left(T_{1}T\right)_{x}^{-1}\left(-\epsilon\right)=\left(T_{1}T\right)_{y}^{-1}\left(\epsilon\right)$
and $\left(T_{1}T\right)_{z}^{-1}\left(-\epsilon\right)=\left(T_{1}T\right)_{z}^{-1}\left(\epsilon\right)$.

\end{widetext}

\section{Fitting the Spin Lattice Relaxation Rate Data}

The expressions for $(T_{1}T)_{\alpha}^{-1}$ given above depend on
six parameters: $\xi_{x}$, $\xi_{y}$, $\xi_{z}$, $\varphi_{xy}$,
$\varphi_{yx}$, and $\varphi_{zz}$. We first fit the zero-strain
data shown in Fig. 3(b) and 3(d) of the main text assuming all the
$\varphi_{\alpha\beta}=0$, and that $\xi_{y}=\xi_{x}$. Because the
Landau damping term, $\Gamma$, is unknown, one cannot simply extract
the $\xi_{x,z}$ directly from the data. However, the ratio of $(T_{1}T)_{x}^{-1}/(T_{1}T)_{z}^{-1}$
does constrain the data and enable us to fit the data using the temperature-dependent
correlation lengths shown in Fig. 4(b) of the main text. The hyperfine
coupling constants are given by: $A_{xx}=A_{yy}=0.66$ T/$\mu_{B}$,
$A_{zz}=0.47$ T/$\mu_{B}$, and $A_{xz}=A_{yz}=0.43$ T/$\mu_{B}$
\cite{takigawa2008}, and we assume the value $A_{xy}=0.33$ T/$\mu_{B}$
\cite{NMRnematicStrainBa122PRB2016}.

Using these values for $\xi_{x,z}$ and assuming that $\xi_{y}=\xi_{x}$,
we then proceed to fit the strain-dependent $(T_{1}T)^{-1}$ data
to the three nematic order parameters, $\varphi_{xy}=\chi_{\mathrm{nem}}^{(xy)}\epsilon$,
$\varphi_{yx}=\chi_{\mathrm{nem}}^{(yx)}\epsilon$, and $\varphi_{zz}=\chi_{\mathrm{nem}}^{(zz)}\epsilon$,
where the $\chi_{\mathrm{nem}}^{(\alpha\beta)}$ are the static nematic
susceptibilities of the three components of the nematic order. These
data are shown in Fig. 4 of the main text as a function of temperature.

\end{document}